\documentclass[aps,twocolumn,superscriptaddress,groupedaddress,showkeys,showpacs]{revtex4-2}
\usepackage{amsmath, amssymb}
\usepackage{amsthm}
\usepackage{graphicx}
\usepackage{physics}
\usepackage{xcolor}
\usepackage{caption}
\usepackage{tikz}
\usetikzlibrary{quantikz2}
\usepackage{float}
\usepackage{amsfonts}
\usepackage{verbatim}
\usepackage{hyperref}
\usepackage{mathcomp}
\usepackage[capitalise, nameinlink]{cleveref}
\usepackage[linesnumbered,ruled,vlined]{algorithm2e}
\usepackage{booktabs}
\usepackage{subcaption}
\usetikzlibrary{trees}
\usetikzlibrary{arrows.meta,positioning,fit,calc,backgrounds,decorations.pathreplacing}
\usetikzlibrary{shapes.geometric, arrows}

\crefname{algocf}{Alg.}{Algs.}
\Crefname{algocf}{Algorithm}{Algorithms}
\Crefname{figure}{Fig.}{Figs.}

\newcommand{\bu}{\boldsymbol{u}}
\newcommand{\bv}{\boldsymbol{v}}
\newcommand{\ba}{\boldsymbol{a}}

\newcommand{\bthe}{{\boldsymbol{\theta}}}
\newcommand{\SOVQD}{{\textbf{SSVQD}}}
\newcommand{\SAVQD}{{\textbf{SAVQD}}}
\newcommand{\hhhh}{$\textbf{H}_4\;$}
\newcommand{\hh}{$\textbf{H}_2\;$}
\newcommand{\lih}{$\textbf{LiH}\;$}

\theoremstyle{plain}

\begin{document}

\title{State-Specific Orbital Optimization for Enhanced 
Excited-States Calculation on Quantum Computers}

\author{Guorui Zhu}
\affiliation{School of Mathematical Sciences, Fudan University, Shanghai 200433, China}
\author{Joel Bierman}
\affiliation{ Department of Physics and Department, Duke University}
\affiliation{Department of Electrical and Computer Engineering, North Carolina State University}
\author{Jianfeng Lu}
\affiliation{Department of Mathematics, Department of Physics and
Department of Chemistry, Duke University}
\email[Corresponding author: ]{jianfeng@math.duke.edu}
\author{Yingzhou Li}
\affiliation{School of Mathematical Sciences, Shanghai Key Laboratory for
Contemporary Applied Mathematics, Fudan University} \affiliation{Key
Laboratory of Computational Physical Sciences, Ministry of Education}
\email[Corresponding author: ]{yingzhouli@fudan.edu.cn}

\begin{abstract}
    We propose a state-specific orbital optimization scheme for improving
    the accuracy of excited states of the electronic structure Hamiltonian
    for the use on near-term quantum computers, which can be combined with
    any overlap-based excited-state quantum eigensolver. We derived the
    gradient of the overlap term between different states generated by
    different orbitals with respect to the orbital rotation matrix and
    use the gradient-based optimization methods to optimize
    the orbitals. This scheme allows for more flexibility in the choice
    of orbitals. We implement the state-specific orbital optimization
    scheme with the variational quantum deflation~(VQD) algorithm, and
    show that it achieves higher accuracy than the state-averaged
    orbital optimization scheme on various molecules including \hhhh and
    \lih.

\end{abstract}

\maketitle

\section{Introduction}
\label{sec:introduction}

One of the most promising applications of quantum computing on near-term
devices is to solve the electronic structure problem---determining the
ground and excited states, along with their corresponding energies,
of an electronic Hamiltonian~\cite{helgaker2013molecular}.
This problem is central to quantum chemistry and materials science,
with applications in areas such as drug discovery, catalysis, and
the development of new materials~\cite{szabo1996modern}.

The general procedure for solving the electronic structure problem on
quantum computers begins with the selection of a finite orbital basis
set. The electronic structure Hamiltonian is then expressed in the
second quantization formalism under this basis. This fermionic Hamiltonian
must be mapped to a qubit Hamiltonian. Once in qubit form, quantum algorithms
most notably the Variational Quantum
Eigensolver~(VQE)~\cite{peruzzo2014variational, tilly2022variational}
or Quantum Phase Estimation~(QPE)~\cite{nielsen2010quantum, kitaev1995qpe}
---are used to estimate the low-lying eigenstates and eigenvalues.
For a quantum algorithm without resource reduction techniques, one qubit
is needed to represent each spin-orbital when using standard encodings 
such as the Jordan--Wigner mapping. Due to the limited number of qubits
on near-term quantum devices, the number of spin-orbitals is limited,
and thus error from the truncation of the orbital set will be
substantial~\cite{kühn2019accuracy, elfving2020will, gonthier2022measurements}.

Similar challenges also arise in classical computational chemistry.
For example, the computational complexity of full configuration
interaction~(FCI) increases combinatorially with the number of orbitals.
To address the issue of a limited number of orbitals, a class of methods
known as orbital optimization has been developed, exemplified by the
complete active space self-consistent field
(CASSCF)~\cite{roos1980casscf, siegbahn1981casscf, siegbahn1980comparison}
and OptOrbFCI~\cite{li2020}. These approaches aim to carefully select
an active orbital subset from a larger orbital space and apply orbital
optimization to improve the accuracy of excited-state calculations.
Such methods have also been extended to the quantum computing domain,
including quantum CASSCF~\cite{yalouz2021saoo, omiya2022saoo} and
OptOrbVQE~\cite{bierman2023, bierman2024}. In these schemes, orbital
optimization is performed by minimizing the weighted average energy
of the states, a strategy known as \emph{state-averaged orbital optimization}.
To maintain completeness while keeping the main text concise, we provide the
basic algorithmic procedure of the state-averaged scheme in \cref{app:state_average}.
Orbital optimization has been shown to have the potential to
achieve higher accuracy than the usual FCI calculation with more
orbitals in our previous works~\cite{bierman2023, bierman2024}.

However, the state-averaged orbital optimization scheme for excited
states encounters some challenges: in general, a single compact set of orbitals
cannot be expected to accurately describe multiple distinct excited
states. High accuracy for an excited state can only be achieved if the
orbitals with significant contributions to the excited state wave 
function are included in the orbital set. If the different
excited states exhibit very distinct wave function component patterns,
then, in the worst case, the total number of orbitals required will
be the sum of the number needed for each individual excited state.
Some further discussions of disadvantages of the state-averaged orbital
optimization can be found in~\cite{marie2023excited, kossoski2023ssci}.

The state-specific orbital optimization scheme is a remedy to the above
mentioned shortcoming of the state-averaged approach, enabling the use and optimization
of orbitals tailored to each individual state. Several studies have
developed state-specific schemes in the classical computational
setting~\cite{marie2023excited, kossoski2023ssci, yalouz2023ss,
saade2024excited}, demonstrating that state-specific orbital
optimization can achieve higher accuracy with even fewer orbitals
compared to the state-averaged case. However, under classical
computation, state-specific orbital optimization is computationally
expensive due to the need for evaluating overlaps between different
 states~\cite{burton2021generalized, burton2022generalized}.
In our previous work~\cite{zhu2025}, we proposed
a quantum algorithm to efficiently compute such overlaps. This enables
the possibility of quantum acceleration for state-specific orbital
optimization.

In this paper, we propose a state-specific orbital optimization scheme
on quantum computers. This scheme can be combined with any overlap-based
excited-state quantum eigensolver.

Our main contributions are as follows:
\begin{itemize}
  \item We introduce a state-specific orbital optimization method, which generalizes
        the state-averaged orbital optimization scheme.
  \item We derive the gradient of the overlap term between states generated by different
        orbitals with respect to the orbital rotation matrix.
  \item We demonstrate how gradient-based optimization methods can be used to optimize
        the orbitals.
  \item We implement the method with the variational quantum deflation (VQD) algorithm
        and show improved accuracy over the state-averaged scheme on various molecules
        such as \hhhh and \lih.
\end{itemize}

The rest of this paper is organized as follows. In
\cref{sec:excited_state_quantum_eigensolvers}, we give a brief review
of excited-state quantum eigensolvers. 
In \cref{sec:state_specific_orbital_optimization}, we present the
state-specific orbital optimization scheme in detail. In
\cref{sec:SOVQD} and \cref{sec:numerical_results}, we demonstrate its
implementation with VQD and show numerical results. At last, we
conclude the paper in \cref{sec:summary}.

\section{Excited-State Quantum Eigensolvers}
\label{sec:excited_state_quantum_eigensolvers}

We begin by reviewing hybrid quantum-classical variational
methods. While the state-averaged scheme places no particular
constraints on the choice of solver, the state-specific scheme, due to its
intrinsic nature, requires overlap-based approaches. This necessity will be
explained in more detail in the following.

Hybrid quantum-classical variational methods have been widely used to
compute the ground state of electronic structure Hamiltonians. The central
idea is to use a parameterized quantum circuit as an ansatz and optimize
its parameters on a classical computer. The objective function, evaluated
on a quantum device, guides the optimization. The variational quantum
eigensolver~(VQE)~\cite{peruzzo2014variational, tilly2022variational}
is a representative example of this approach.
Here we focus on extensions of VQE for computing excited states. These
methods can be divided into two categories: subspace methods and overlap-based methods.

Subspace methods include the multi-configurational variational quantum
eigensolver~(MCVQE)~\cite{parrish2019mcvqe} and the subspace-search
VQE~(SSVQE)~\cite{nakanishi2019ssvqe}. These methods construct a set
of mutually orthogonal states and apply the same ansatz of quantum circuit to each state. The
objective is to minimize the trace or weighted trace of the
Hamiltonian projected onto the subspace spanned by these states.
More specifically, given a set of orthogonal states
$\{\ket{\Psi_\alpha}\}_{\alpha=1}^K$,
the subspace methods minimize the following objective function:
\begin{equation}\label{eq:subspace_objective_function}
    F(\bthe) = \sum_{\alpha=1}^K w_\alpha \bra{\Psi_\alpha}
    \Theta(\bthe)^\dag\mathcal{H}\Theta(\bthe)\ket{\Psi_\alpha},
\end{equation}
where $w_\alpha$ are the weights for each state, $\mathcal{H}$ is the
Hamiltonian of the system and $\Theta(\bthe)$ is the ansatz circuit.
The orthogonality of the states is guaranteed since the
ansatz circuit $\Theta(\bthe)$ is unitary. These methods
are not compatible with state-specific orbitals. The reason
is that when different basis is used, orthogonality between
the resulting states is not guaranteed. More specifically, for two
basis sets $\{\psi^{(\alpha)}_i\}_{i=1}^N$ and
$\{\psi^{(\beta)}_i\}_{j=1}^N$, the overlap between the states $\Theta(\theta)\ket{\Psi_\alpha}$ and
$\Theta(\theta)\ket{\Psi_\beta}$ is given by
\begin{equation*}    
\bra{\Psi_\alpha}\Theta^\dag(\theta)
U\bigl( \langle \psi^{(\alpha)} \mid \psi^{(\beta)}\rangle \bigr) \Theta(\theta)\ket{\Psi_\beta},
\end{equation*}
where $\bra{\psi^{(\alpha)}}\ket{\psi^{(\beta)}}$ is a
matrix whose $ij-$th element equals to $\langle \psi_i^{(\alpha)} \mid \psi_j^{(\beta)} \rangle$ and
$U\bigl(\bra{\psi^{(\alpha)}}\ket{\psi^{(\beta)}}\bigr)$ is the non-unitary orbital transformation
as discussed in \cite{zhu2025}.
In general, this is not equal to
\begin{equation*}
\bra{\Psi_\alpha}U\bigl( \langle \psi^{(\alpha)} \mid \psi^{(\beta)}\rangle \bigr)\ket{\Psi_\beta}.
\end{equation*}

The overlap-based methods, on the other hand, use explicit penalty term
to enforce orthogonality between states.
While also minimizing the trace or weighted trace of the Hamiltonian,
they introduce a penalty term to discourage overlap between states.
Therefore, state-specific
orbitals can be used, but at the cost of requiring the computation of
many pairwise overlaps. Representative examples of the latter include
the variational quantum  deflation~(VQD)~\cite{higgott2019variational}
and the quantum orbital minimization method~(qOMM)~\cite{bierman2022qomm}.

Our work builds upon overlap-based methods, with a particular focus
on the VQD algorithm. 
VQD solve the excited-state by projecting the excited-state wave
function into the subspace orthogonal to the lower states.
The algorithm is as follows:
\begin{enumerate}
    \item Solve the ground state by VQE, which gives the ground state
          wave function $\ket{\Psi_1}$.
    \item Suppose we are solving for the $k$-th excited state $\ket{\Psi_{k+1}}$, 
          and that the ground state as well as all previously obtained excited states 
          $\ket{\Psi_j}$ for $1 \leq j \leq k$ have already been determined. 
          Then we can construct the deflated Hamiltonian
          \begin{equation*}
              \mathcal{H}_{k+1} = \mathcal H + \sum_{j=1}^{k} \beta_j
              \ket{\Psi_j}\bra{\Psi_j},
          \end{equation*}
          where $\mathcal H$ is the Hamiltonian of the system. The $k$-th
          excited state can be obtained by the ground state of the deflated
          Hamiltonian $\mathcal{H}_{k+1}$ via VQE, as long as $\beta_j$ are chosen
          to satisfy the following condition
          \begin{equation*}
              \beta_j > E_{k+1} - E_j, \quad \forall j = 1, 2, \dots, k.
          \end{equation*}
    \item Repeat the above step until all excited states needed are obtained.

\end{enumerate}
The ground state of deflated Hamiltonian $\mathcal{H}_{k+1}$ can be solved
by variational principle, i.e., minimizing the expectation value
\begin{equation}\label{eq:VQD_expectation_value}
    \begin{split}
        \Psi_{k+1} &=  \arg\min_{\Psi} \bra{\Psi} \mathcal{H}_{k+1} \ket{\Psi} \\
        &= \arg\min_{\Psi} \left\{ \bra{\Psi} \mathcal{H} \ket{\Psi} +
        \sum_{j=1}^{k} \beta_j \abs{\bra{\Psi_j}\ket{\Psi}}^2 \right\}.
    \end{split}
\end{equation}

\section{State-Specific Orbital Optimization}
\label{sec:state_specific_orbital_optimization}

The state-specific orbital optimization scheme is an improvement of the
state-averaged orbital optimization scheme, which can use and optimize
the specific orbitals for each state. 

First, we introduce the notation for the state-specific orbitals and give
a general form of the objective function. Suppose
that we want to solve the low-lying $K$ states of a system. For the
$k$-th state, we introduce a set of orbitals $\{\psi_j^{(k)}\}_{j=1}^N$,
which are defined by rotating a given basis set $\{\phi_i\}_{i=1}^M$ with
an $M$ times $N$ partial unitary matrix $\bu^{(k)}$ as follows:
\begin{equation*}
    \psi_j^{(k)} = \sum_{i=1}^M \phi_i \bu^{(k)}_{ij}, \quad j=1,2,\dots,N.
\end{equation*}

\begin{figure}[htb]
    \centering
    \includegraphics[width=0.3\textwidth]{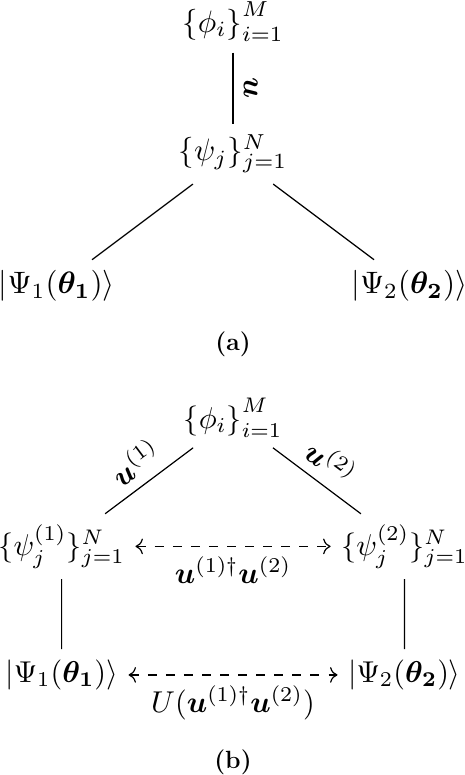}
    \caption{Comparison between (a) state-averaged and (b) state-specific orbital optimization.}
    \label{fig:orbital_opt}
\end{figure}

A comparison of the state-averaged and state-specific orbitals
can be found in \cref{fig:orbital_opt}.  As mentioned in
\cref{sec:excited_state_quantum_eigensolvers}, since different orbitals
are used for different states, we will carry out orbital optimization with the
overlap-based methods, such as VQD and qOMM. For the overlap-based methods,
the general objective function $\textbf{F}_{\text{SS}}$ for the state-specific
orbital optimization can be expressed as
\begin{equation*}
    \textbf{F}_{\text{SS}}(\{\bthe_k, \bu^{(k)}\}_{k=1}^{K}) =
          \textbf{G}\left(\{\textbf{E}_{\text{SS}}^k\}_{k=1}^{K},
         \{\textbf{O}_{\text{SS}}^{jk}\}_{1\le j<k\le K}\right),
\end{equation*}
where
\begin{equation}\label{eq:state_specific_energy_and_overlap}
    \begin{split}
        \ket{\Psi_k(\bthe_k, \bu^{(k)})} &= \Theta_k(\bthe_k)\ket{\Psi_k;\bu^{(k)}}, \\
        \textbf{E}_{\text{SS}}^k(\bthe_k, \bu^{(k)}) =
         & \bra{\Psi_k(\bthe_k, \bu^{(k)})}
            \mathcal{H}\ket{\Psi_k(\bthe_k, \bu^{(k)})},\\
        \textbf{O}_{\text{SS}}^{jk}(\bthe_j, \bthe_k, \bu^{(j)}, \bu^{(k)}) =
         & \left|\bra{\Psi_j(\bthe_j, \bu^{(j)})}
        \ket{\Psi_k(\bthe_k, \bu^{(k)})}\right|^2,
    \end{split}
\end{equation}
and $\textbf{G}$ is a function that combines the energy and overlap terms.
Here $\ket{\Psi_k;\bu^{(k)}}$ is a reference state defined under the
basis set $\{\psi_j^{(k)}\}_{j=1}^N$, and $\Theta_k(\bthe_k)$ is the
ansatz circuit for the $k$-th state with parameters $\bthe_k$.

The energy expectation $\textbf{E}_{\text{SS}}^k$ can be simplified such that
the dependence on $\bu^{(k)}$ is explicit.
If we take the Fermionic second-quantized Hamiltonian as the following form:
\begin{equation}
    \begin{split}
        \mathcal{H} =& \frac{1}{2}\sum_{i,j=1}^M h_{ij} \ba(\phi_i)^\dagger \ba(\phi_j) \\ 
         & + \frac{1}{4}\sum_{i,j,k,l=1}^M v_{ijkl} \ba(\phi_i)^\dagger
         \ba(\phi_j)^\dagger \ba(\phi_k) \ba(\phi_l),
    \end{split}
\end{equation}
where $h_{pq}$ and $v_{pqrs}$ are the one- and two-electron integrals
in the basis set $\{\phi_i\}_{i=1}^M$, $\ba(\phi_i)$ and $\ba(\phi_i)^\dagger$ are
the annihilator and creator operator with respect to the state $\phi_i$,
then this orbital rotation is equivalent to transforming the Hamiltonian as
\begin{equation}
    \begin{split}
        \tilde{\mathcal{H}}(\bu) =&  \sum_{p,q=1}^N\sum_{i,j=1}^M h_{ij}
        \bu_{ip}\bu_{jq}\cdot
        \ba(\psi_p)^\dagger \ba(\psi_q) \\
        &+ \frac{1}{2}\sum_{p,q,r,s=1}^N\sum_{i,j,k,l=1}^M
        v_{pqrs}\bu_{ip}\bu_{jq}\bu_{kr}\bu_{ls} \\
        &\qquad\qquad\qquad\qquad
        \cdot\ba(\psi_p)^\dagger 
        \ba(\psi_q)^\dagger \ba(\psi_s) \ba(\psi_r).
    \end{split}
\end{equation}
Under this transformation, the $\textbf{E}_{\text{SS}}^k$ can be expressed as
\begin{equation}\label{eq:state_specific_energy}
    \begin{split}
        \textbf{E}_{\text{SS}}^k = &
        \sum_{p,q=1}^N\sum_{i,j=1}^M h_{ij}\bu^{(k)}_{ip}\bu^{(k)}_{jq} 
        \cdot\mathcal{R}^p_{q}(\Psi_k(\bthe_k)) \\
        &+ \frac{1}{2}\sum_{p,q,r,s=1}^N\sum_{i,j,k,l=1}^M
        v_{pqrs}\bu^{(k)}_{ip}\bu^{(k)}_{jq}\bu^{(k)}_{kr}\bu^{(k)}_{ls} \\
        &\qquad\qquad\qquad\qquad\qquad\quad \cdot\mathcal{R}^{pq}_{rs}(\Psi_k(\bthe_k)),
    \end{split}
\end{equation}
where
\begin{equation}
    \begin{split}
        \mathcal{R}^p_{q}(\Psi_k(\bthe_k)) &= \bra{\Psi_k(\bthe_k)} \ba(\psi_p)^\dagger \ba(\psi_q) 
        \ket{\Psi_k(\bthe_k)}, \\
        \mathcal{R}^{pq}_{rs}(\Psi_k(\bthe_k)) &= \\
        \bra{\Psi_k(\bthe_k)}& \ba(\psi_p)^\dagger 
        \ba(\psi_q)^\dagger \ba(\psi_s) \ba(\psi_r) \ket{\Psi_k(\bthe_k)}
    \end{split}
\end{equation}
are the one- and two-electron reduced density matrices
(RDMs) of the state $\ket{\Psi_\alpha}$ in the basis set $\{\psi_j\}_{j=1}^N$.

This type of objective function has been studied in the state-specific
orbital optimization proposed before in~\cite{marie2023excited,
kossoski2023ssci, yalouz2023ss, saade2024excited}. However, they faced the
problem that calculating the overlap between different states is
computationally expensive, especially when the number of orbitals is large.
Therefore, we divide the challenge into two main tasks: The first is to
efficiently compute the overlap between states generated by different orbitals,
and the second is to carry out orbital optimization using these overlap terms.

Our previous work~\cite{zhu2025}
proposed a quantum algorithm to accomplish the first task, which costs an external
circuit of depth $O(N)$, where $N$ is the number of orbitals. This gives us
an opportunity to implement the state-specific orbital optimization scheme
on a quantum computer. In the notation of~\cite{zhu2025}, 
if $\bu^{(1)}$ and $\bu^{(2)}$ are two
orbital rotation matrices, then the overlap between the states $\ket{\Psi_1}$
and $\ket{\Psi_2}$ generated by the orbitals $\bu^{(1)}$ and $\bu^{(2)}$ can be
expressed as
\begin{equation}\label{eq:overlap_between_states_rotation}
    \bra{\Psi_1}U(\bu^{(1)\top}\bu^{(2)})\ket{\Psi_2},
\end{equation}
where $U(\bu^{(1)\top}\bu^{(2)})$ is the non-unitary orbital transformation.
The non-unitary transformation $U(\bu^{(1)\top}\bu^{(2)})$ can be implemented
as a block-encoding of an $O(N)$ depth quantum circuit with about $2N$ qubits.

For the second task, i.e, the optimization of the orbitals, one can try to use
the constrained derivative-free optimization methods~\cite{powell2007view},
such as COBYLA~\cite{powell1994cobyla}, SLSQP~\cite{kraft1988slsqp} or
UPOQA~\cite{liu2025upoqa}, to optimize
the orbitals. But these methods will try to convert the constrained optimization
to to unconstrained optimization by Lagrange multipliers, which will lead to
poor convergence behavior when the dimension of the optimization problem is high.

Here, we adopt gradient-based optimization methods to optimize the orbitals. To
this end, we require the gradient of the objective function $\textbf{F}_{\text{SS}}$
with respect to the orbital rotation matrices ${\{\bu^{(k)}\}}_{k=1}^{K}$. The gradient
of the energy term $\textbf{E}_{\text{SS}}^k$ with respect to $\bu^{(k)}$ can be
readily obtained from \cref{eq:state_specific_energy}, since its dependence on
$\bu^{(k)}$ is explicit. Importantly, the gradient of the energy expectation term
only involves $1$- and $2$-RDMs, which can be computed with $O(N^4)$ quantum circuits
and reused throughout the optimization of $\bu^{(k)}$. The gradient of the overlap
term, whose derivation is more involved, will be presented in
\cref{app:gradient_of_overlap}. The result is shown in \cref{eq:gradient_of_overlap}.
\begin{equation}\label{eq:gradient_of_overlap}
    \begin{split}
        \frac{\partial \textbf{O}_{\text{SS}}^{12}}{\partial \bu^{(2)}_{pq}} 
        = &\frac{\partial}{\partial \bu^{(2)}_{pq}}|\bra{\Psi_1}
        U(\bu^{(1)\top}\bu^{(2)})\ket{\Psi_2}|^2 \\
        = & \bra{\Psi_2}U(\bu^{(2)\top}\bu^{(1)}) \ket{\Psi_1} \\
          &\cdot \sum_{i=1}^N
          \bu^{(1)}_{pi}
          \bra{\Psi_1} \ba_i^\dag U(\bu^{(1)\top}\bu^{(2)})\ba_q
          \ket{\Psi_2} + \text{c.c.}.\\
    \end{split}
\end{equation}
This gradient requires $O(N^2)$ quantum circuits to compute $1$-RDM-like terms,
each time the rotation matrix $\bu^{(1)}$ or $\bu^{(2)}$
is updated, we need to re-construct the circuit and re-evaluate the gradient.

One can also take higher order derivatives of the objective function. For the
energy expectation term, once the RDMs are computed, we can easily obtain the
higher order derivatives by classical computation. And thanks to the explicit
form of the gradient of $U(\bu^{(i)\top}\bu^{(j)})$, the higher order derivatives
of the overlap term can also be obtained. Thus higher order optimization methods
can be used to optimize the orbitals. But every time we take the derivative for
the overlap term, the cost will increase by a factor of $N^2$. For example, the
second order derivative of the overlap term needs to compute all terms like
\begin{equation*}
    \bra{\Psi_1} \ba_i^\dag\ba_j^\dag U(\bu^{(1)\top}\bu^{(2)})
    \ba_k\ba_l\ket{\Psi_2}.
\end{equation*}
One should always consider the trade-off between the cost of computing and the
convergence of the optimization methods.

Given the gradient of the overlap term, appropriate overlap-based excited-state
solver can be used for the state-specific
orbital optimization. In the next section, we will use the VQD algorithm and only
first order optimization methods to optimize the orbitals.

\section{State-Specific
Orbital Optimization VQD~(\SOVQD)}
\label{sec:SOVQD}

In this section, we present the state-specific orbital optimization VQD~(\SOVQD)
algorithm. VQD is chosen as the base algorithm due to its simplicity in both
understanding and implementation, which makes it well suited to demonstrate
how the proposed state-specific orbital optimization scheme can be implemented.

As discussed in \cref{sec:state_specific_orbital_optimization}, for the $k$-th
state we introduce an $M \times N$ partial unitary matrix $\bu^{(k)}$ to perform
a basis rotation. In the rotated basis defined by $\bu^{(k)}$, the $k$-th
state is represented by a parameterized quantum circuit $\ket{\Psi(\bthe_k)}$,
where $\bthe_k$ denotes the parameters of the ansatz circuit.
The objective function $\textbf{F}_k(\bthe_k, \bu^{(k)})$ for the $k$-th excited
state is defined as
\begin{equation}
        \textbf{F}_k(\bthe_k, \bu^{(k)})
        = \textbf{E}_{\text{SS}}^k(\bthe_k, \bu^{(k)}) +
        \sum_{j=1}^{k-1} \beta_j \textbf{O}_{\text{SS}}^{jk}(\bthe_k, \bu^{(k)}).
\end{equation}
Here we have assumed that the lower $k-1$ states
$\{\ket{\Psi_j}\}_{j=1}^{k-1}$ and their corresponding orbital rotations
$\{\bu^{(j)}\}_{j=1}^{k-1}$ have already been obtained. Thus we can treat
them as constants in the optimization of the $k$-th state and omit the dependence
of $\textbf{F}_k$ and $\textbf{O}_{\text{SS}}^{jk}$ on these variables for simplicity. 

\begin{figure}[htb]
    \centering
    \includegraphics[width=0.45\textwidth]{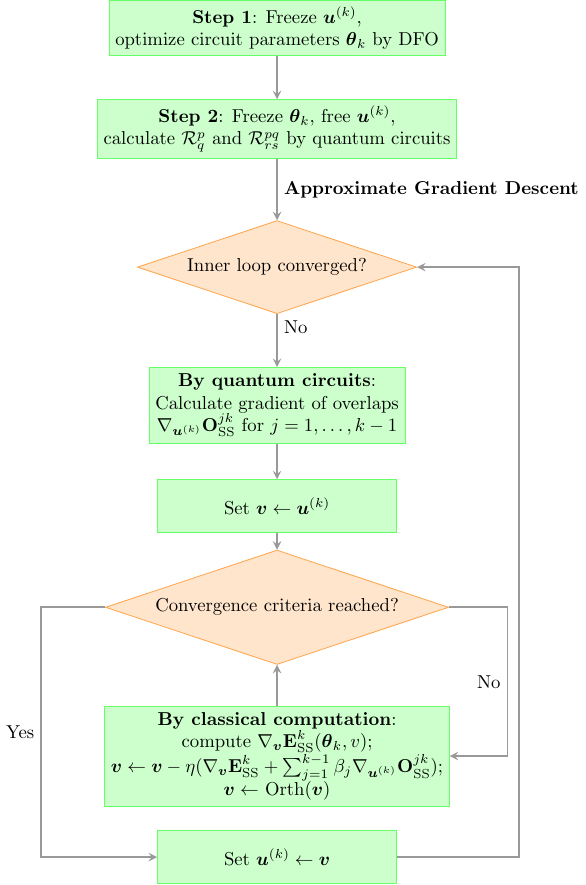}
    \caption{The workflow of the optimization approach.}
    \label{fig:two_step_optimization}
\end{figure}

Although nonlinear derivative-free optimization~(DFO) methods can be applied,
their convergence is often poor due to the high dimensionality and constraints
of the optimization problem. Hence, in this work we employ DFO methods solely
for optimizing the circuit parameters $\bthe_k$, while the orbital parameters
$\bu^{(k)}$ are optimized using gradient-based methods. The iterative procedure
for the $k$-th state alternates between two steps:
\begin{enumerate}
\item First, optimize the parameters $\bthe_k$ with the orbitals $\bu^{(k)}$
      held fixed, employing DFO methods.
\item Next, with the optimized $\bthe_k$ fixed, optimize orbital rotation
      $\bu^{(k)}$ by gradient-based methods.
\end{enumerate}
Throughout the following discussion, this procedure will be referred to as a
single two-step iteration. \cref{fig:two_step_optimization} shows the
workflow of this optimization procedure. 

We now proceed to describe the second step of the procedure shown in
\cref{fig:two_step_optimization} in detail. In the second step, the gradient of the
objective function $\textbf{F}_k(\bthe_k, \bu^{(k)})$ with respect to the orbital
rotation $\bu^{(k)}$ can be calculated separately by two parts.

The first part is the gradient of the $\textbf{E}_{\text{SS}}^k$ term, which is
defined in \cref{eq:state_specific_energy_and_overlap}. This term is a degree-$4$
polynomial in the orbital parameters $\bu^{(k)}$ by \cref{eq:state_specific_energy},
and its gradient can be obtained via the chain rule. Once these RDMs are available,
all gradients of the expectation value can be efficiently evaluated on a classical
computer.

The second part is the gradient of the overlap terms $\textbf{O}_{\text{SS}}^{jk}$ which is
also defined in \cref{eq:state_specific_energy_and_overlap}. By \cref{eq:gradient_of_overlap},
the gradient of the overlap term can be expressed as
\begin{equation}\label{eq:gradient_of_rotation}
    \begin{split}
        \frac{\partial \textbf{O}_{\text{SS}}^{jk}}{\partial \bu^{(k)}_{pq}} 
        = & \frac{\partial}{\partial \bu^{(k)}_{pq}} 
        \abs{\bra{\Psi_j}U(\bu^{(j)\top}\bu^{(k)})\ket{\Psi(\bthe_k)}}^2 \\
        = & \bra{\Psi(\bthe_k)}U(\bu^{(k)\top}\bu^{(j)}) \ket{\Psi_j} \\
          &\cdot \big(\sum_{l=1}^N 
          \bu^{(j)}_{pl}\bra{\Psi_j}\ba^\dagger_l 
          U(\bu^{(j)\top}\bu^{(k)})a_q\ket{\Psi(\bthe_k)}\big) \\ &+ \text{c.c.}.
    \end{split}
\end{equation}

Compared with the gradient of the $\mathbf{E}_{\text{SS}}^k$ term, the gradient
of the overlap $\textbf{O}_{\text{SS}}^{jk}$ contribution must be re-evaluated after
each orbital rotation update, which requires $O(N^2)$ quantum circuits. Nevertheless,
this overhead is still significantly lower than the cost of evaluating the full
objective function, which involves $O(N^4)$ quantum circuits to compute all one- and
two-RDMs. Therefore, if the number of gradient descent steps is modest, the additional
cost of evaluating the overlap gradient is negligible compared with that of
derivative-free optimization of the ansatz parameters $\bthe_k$. We note, however,
that unlike in conventional CASSCF where second-order methods are typically employed
for orbital optimization, here such approaches would be prohibitively expensive, since
every orbital update would require recomputing all $O(N^4)$ quantum circuits. This
motivates our choice of gradient-based first-order optimization methods.

Since the gradient of the expectation value term can be calculated with
high efficiency, one strategy is to freeze the gradient of the overlap term
and only calculate the gradient of the expectation value term
every gradient descent step. Then after $L$ steps, update the
gradient of the overlap term. This can reduce the cost of the
optimization significantly. In our numerical experiments, $L$
is set to $100$. The pseudo-code of this gradient descent
optimization is shown in \cref{alg:gradient_descent_SOVQD}.
\begin{algorithm}[h]
    \caption{Orbital optimization in \SOVQD}
    \label{alg:gradient_descent_SOVQD}
    \KwIn{$\mathcal{H}$, $\{\Psi_j\}_{j=1}^{k-1}$, $\{\bu^{(j)}\}_{j=1}^{k-1}$,
    $\bthe_k$, $L$, $\eta$, $\bu^{(k)}_{\text{init}}$, $\{\beta_j\}_{j=1}^{k-1}$}
    \KwOut{$\bu^{(k)}$}
    Initialize $\bu^{(k)} = \bu^{(k)}_{\text{init}}$\;
    Calculate the $\mathcal{R}^p_{q}(\Psi(\bthe_k))$ and
    $\mathcal{R}^{pq}_{rs}(\Psi(\bthe_k))$ by quantum circuits\;
    \While{not converged}{
        Calculate the gradient of the overlap terms
        $\nabla_{\bu^{(k)}}\textbf{O}_{\text{SS}}^{jk}(\bthe_k,\bu^{(k)})$
        for all $j = 1, 2, \dots, k-1$
        by \cref{eq:gradient_of_rotation}\;
        $\bv \gets \bu^{(k)}$, $\l \gets 0$\;
            \If{not converged and $\l \le L$}{
                $\l \gets \l + 1$\;
                Calculate $\nabla_{\bv}\textbf{E}_{\text{SS}}^k(\bthe_k,\bv)$
                on classical computer\;
                $\bv \gets \bv - \eta (\nabla_{\bv}\textbf{E}_{\text{SS}}^k(\bthe_k,\bv)
                - \sum_{j=1}^{k-1}\beta_j\nabla_{\bu^{(k)}}
                \textbf{O}_{\text{SS}}^{jk}(\bthe_k,\bu^{(k)}))$\;
                $\bv \gets \text{Orth}(\bv)$\;
            }
            $\bu^{(k)} \gets \bv$ \;
    }
    Return $\bu^{(k)}$\;
\end{algorithm}

In the end, the psedo-code of computing $k$-th state
with the state-specific orbital optimization VQD algorithm is summarized in
\cref{alg:k_th_SOVQD}.

\begin{algorithm}[h]
    \caption{$k$-th state with \SOVQD}
    \label{alg:k_th_SOVQD}
    \KwIn{$\mathcal{H}$, $\{\Psi_j\}_{j=1}^{k-1}$, 
    $\{\bu^{(j)}\}_{j=1}^{k-1}$,  $L$, $\eta$,
    $\bu^{(k)}_{\text{init}}$, $\bthe_{k\;{\text{init}}}$, $\{\beta_j\}_{j=1}^{k-1}$}
    \KwOut{$\Psi_k$, $\bu^{(k)}$}
    Initialize $\bu^{(k)} = \bu^{(k)}_{\text{init}}$, $\bthe_k = 
    \bthe_{k\;\text{init}}$ \;
    \While{not converged}{
        Freeze $\bu^{(k)}$ and free $\bthe_k$, optimize $\bthe_k$ in
        $\textbf{F}_k(\bthe_k, \bu^{(k)})$ by specific methods\;
        Freeze $\bthe_k$ and free $\bu^{(k)}$, call
        \cref{alg:gradient_descent_SOVQD} to optimize orbital rotation
        $\bu^{(k)}$\;
    }
    Return $\Psi_k = \ket{\Psi(\bthe_k)}$, $\bu^{(k)}$\;
\end{algorithm}

\section{Numerical Results}
\label{sec:numerical_results}

In this section, we will present the numerical results of the
state-specific orbital optimization VQD algorithm on small molecules.
They are compared with the state-averaged orbital optimization VQD~(\SAVQD)
algorithm. Here we emphasize that the overlap terms in \SOVQD\;are
calculated by the quantum circuits in \cite{zhu2025}, which needs
about twice the number of qubits since the states are in different
basis sets. So it's challenging for classical computers to simulate the
\SOVQD\;algorithm. Overall we demonstrate that the \SOVQD\;algorithm
can achieve more accurate results than \SAVQD\;in all cases we have
tested.

To reduce the redundancy, the following settings are assumed to be
identical across all experiments. To preserve spin symmetry, the
partial unitary matrix $\bu$ is chosen as a block-diagonal matrix
with two identical blocks: one for the $\alpha$-spin orbitals and
one for the $\beta$-spin orbitals. The size of $\bu$ is $M \times N$,
where $M$ is the total number of spin-orbitals and $N$ is the number
of active spin-orbitals. The step size $\eta$ in the gradient
descent optimization of the orbital rotation is fixed at $10^{-3}$, with the
update of the overlap gradient performed every $L=100$ steps. 
The parameter $\beta_j$ in the deflated Hamiltonian is set to 
$15\,\text{Ha}$ for all tests. We employ the UCCSD ansatz with 
two repetitions. For both \SAVQD\;and \SOVQD, the ansatz circuits 
are initialized with all parameters set to zero, and the initial 
reference state is taken to be the Hartree--Fock ground state.
The states are indexed starting from $1$, with state~$1$ corresponding to
the ground state and higher indices denoting higher excitation levels. 
Degenerate states with the same energy are distinguished by different indices.

\subsection{\hh}

We begin with our results for the simplest model tested, the ground state
and the first excited state energies of \hh at the nearequilibrium bond
distance of
$0.735$ $\mathring{A}$. We use $6$-$31$g ($4$ orbitals, i.e., $8$
spin-orbitals) as the starting basis and an active space of $4$
optimized spin-orbitals is used. All of the orbital rotations
are initialized with the padded identity matrix, i.e.,
\begin{equation*}
    \begin{pmatrix}
        1 & 0 \\
        0 & 1 \\
        0 & 0 \\
        0 & 0
    \end{pmatrix}
\end{equation*}
It should be noted that in \SAVQD, a single partial unitary
matrix $\bu$ is shared by all states, whereas in \SOVQD\;each
state has its own partial unitary matrix $\bu^{(k)}$, although
they are initialized identically. The results are
shown in \cref{tab:hh_results}. Since for this small system the 
convergence of the both algorithms is fast, we only display the
converged results in the table. The criterion for convergence is that
the energy difference between the current and previous two-step iterations
is less than $10^{-4}$ Hartree.      
The overlap terms $\abs{\bra{\Psi_j}U(\bu^{(j)\top}
\bu^{(k)})\ket{\Psi_k}}^2$ are less than $10^{-8}$ at the end of the
optimization.

Another meaningful result is the weighted energy sum, since the
SAVQD is designed to optimize the orbitals to minimize this value of
the system. The weighted energy sum is also shown in \cref{tab:hh_results}.
In this table, the weight is $2,1$ for the state $1,2$.
The \SOVQD\;gives a better result than the \SAVQD\;even in this case.

\begin{table}[htb]
    \centering
    \begin{tabular}{lccc}
        \toprule
        \textbf{Method} & \textbf{Level 1} & \textbf{Level 2} & \textbf{Weighted sum} \\
        \midrule
        \multicolumn{4}{c}{\textit{Relative error} $\frac{E-E_{\text{FCI}}}{|E_{\text{FCI}}|}$} \\
        \midrule
        \SAVQD & $7.8\times 10^{-3}$ & $5.3\times 10^{-3}$ & $7.1\times 10^{-3}$\\
        \SOVQD & $2.9\times 10^{-3}$ & $1.0\times 10^{-3}$ & $2.3\times 10^{-3}$\\
        \midrule
        \multicolumn{4}{c}{\textit{Energy (a.u.)}} \\
        \midrule
        \textbf{HF} & $-1.847$ & $-1.443$ & $-5.136$\\
        \SAVQD & $-1.857$ & $-1.466$ & $-5.180$\\
        \SOVQD & $-1.866$ & $-1.472$ & $-5.205$\\
        \textbf{FCI} & $-1.872$ & $-1.474$ & $-5.217$\\
        \bottomrule
    \end{tabular}
    \caption{
        Results of \SOVQD\;and \SAVQD\;on \hh (basis set: 6-31g) for the first $2$
        low-lying energies. The upper block shows the relative error
        $\tfrac{E-E_{\text{FCI}}}{|E_{\text{FCI}}|}$ with respect to 6-31g FCI,
        and the lower block gives the absolute energies.
    }
    \label{tab:hh_results}
\end{table}

\subsection{\hhhh}

Now we present the results for the low-lying $5$ energy levels of \hhhh molecule,
a toy system composed of four hydrogen atoms arranged in a square with a
nearest-neighbor distance of $1.23$ $\mathring{A}$. The starting basis set is
cc-pVDZ ($20$ orbitals, i.e., $40$ spin-orbitals), and an active space of $8$
optimized spin-orbitals is used. All of the orbital rotations are initialized
with the padded identity matrix, i.e.,
\begin{equation*}
    \begin{pmatrix}
        1 & 0 & 0 & 0\\
        0 & 1 & 0 & 0\\
        0 & 0 & 1 & 0\\
        0 & 0 & 0 & 1\\
        0 & 0 & 0 & 0\\
        \vdots & \vdots & \vdots & \vdots\\
        0 & 0 & 0 & 0\\
    \end{pmatrix}.
\end{equation*}
All settings for the \SAVQD\;and \SOVQD\;are again, in \SAVQD\;a single
partial unitary matrix $\bu$ is shared by all states, whereas in \SOVQD\;each
state has its own matrix $\bu^{(k)}$, although they are initialized identically.

\begin{figure}
    \centering
    \includegraphics[height=0.9\textheight]{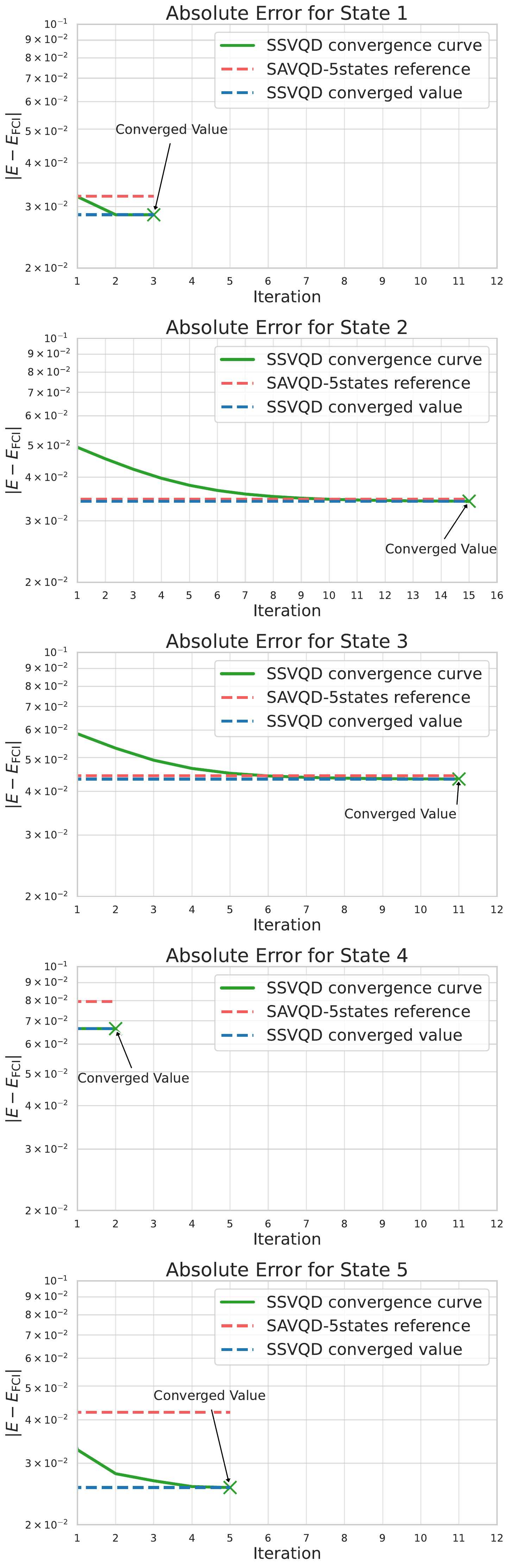}
    \caption{
        The results of \SOVQD\;and \SAVQD\;on \hhhh for the low-lying $5$
        eigen-energies. The $x$-axis denotes the number of two-step iterations,
        while the $y$-axis shows the absolute value of the energy difference
        between the cc-pVDZ FCI energy and the calculated energy.
    }
    \label{fig:hhhh_results}
\end{figure}

The results are shown in \cref{fig:hhhh_results}.
The overlap terms $\abs{\bra{\Psi_j}U(\bu^{(j)\top}\bu^{(k)})\ket{\Psi_k}}^2$
are less than $10^{-8}$ at the end of each two-step iteration. We present the
energy convergence curves and terminate the optimization once the convergence
criterion is satisfied. Specifically, convergence is defined as the energy
difference between two consecutive two-step iterations falling below
$3 \times 10^{-5}$ Hartree. It is important to note that the state-averaged
two-step iteration is shared across all states, whereas the state-specific two-step
iterations are performed independently for each state. Therefore, we include
the state-averaged result as a straight reference line for comparison.

Another meaningful result is the weighted energy sum, since the
SAVQD is designed to optimize the orbitals to minimize this value of
the system. The weighted energy sum is shown in \cref{tab:hhhh_total_energy}.
In this table, the weight is $5,4,3,2,1$ for the state $1,2,3,4,5$.
The \SOVQD\;gives a better result than the \SAVQD\; even in this case.

\begin{table}[ht]
    \centering
    \begin{tabular}{cccccc}
        \toprule
        State & Method & \textbf{HF} & \SAVQD & \SOVQD & \textbf{FCI} \\
        \midrule
        1 & $E$ (a.u.) & $-4.345$ & $-4.398$ & $-4.402$ & $-4.430$ \\
        2 & $E$ (a.u.) & $-4.318$ & $-4.392$ & $-4.392$ & $-4.427$ \\
        3 & $E$ (a.u.) & $-4.261$ & $-4.305$ & $-4.306$ & $-4.349$ \\
        4 & $E$ (a.u.) & $-4.232$ & $-4.256$ & $-4.268$ & $-4.334$ \\
        5 & $E$ (a.u.) & $-3.993$ & $-4.179$ & $-4.193$ & $-4.221$ \\
        \midrule
        Weighted sum & $E$ (a.u.) & $-64.223$ & $-65.160$ & $-65.223$ & $-65.793$ \\
        \bottomrule
    \end{tabular}
    \caption{
        Energies of the first $5$ low-lying states of the \hhhh molecule 
        in the cc-pVDZ basis set, calculated using \SOVQD\;and \SAVQD. 
        Here the values used for the \SAVQD\;are first $5$ low-lying states.
        The last row shows the weighted energy sum. We also left the
        Hartree--Fock energy here for reference.
    }
    \label{tab:hhhh_total_energy}
\end{table}

\subsection{\lih}
\label{sec:lih}

\begin{figure}
    \centering
    \includegraphics[height=0.72\textheight]{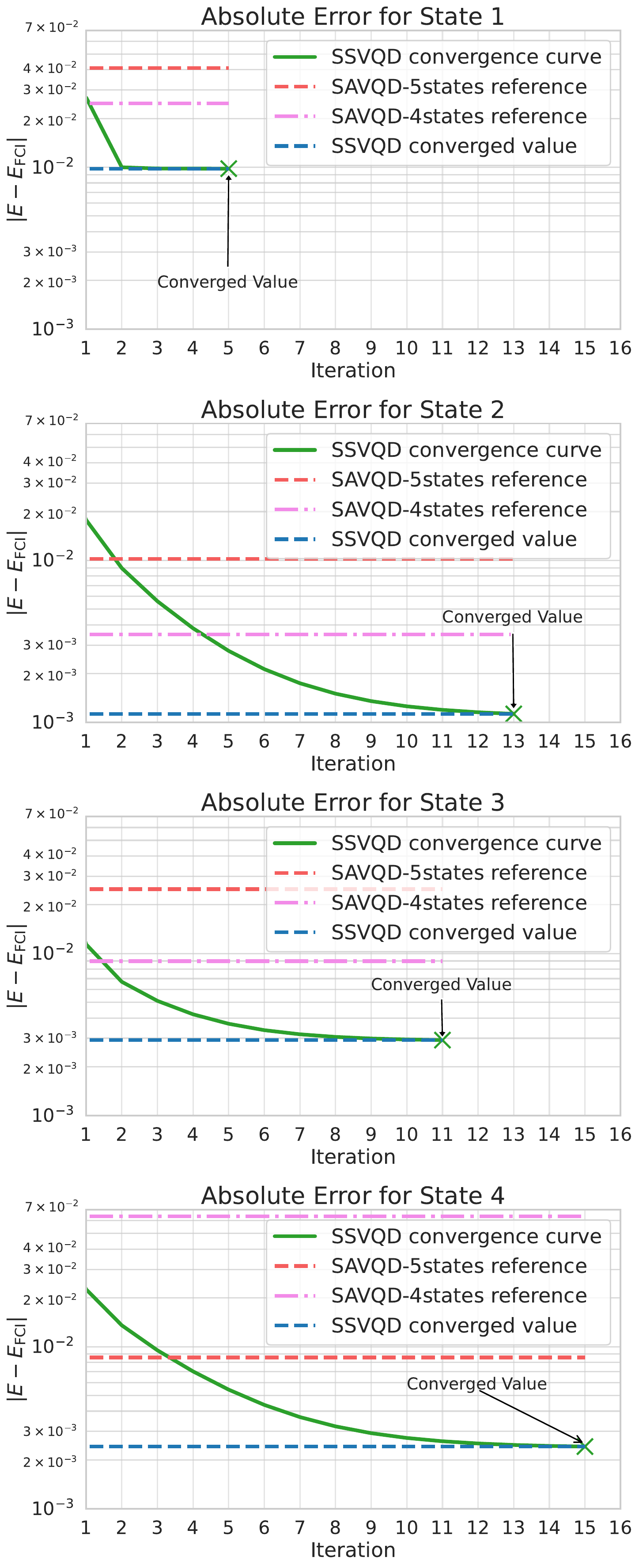}
    \caption{
        The results of \SOVQD\;and \SAVQD\;on \lih for the low lying $4$ and $5$ eigen-energies. 
        The $x$-axis denotes the number of two-step iterations, while the $y$-axis shows the
        absolute value of the energy difference between the cc-pVDZ FCI energy and the calculated energy. 
        Only the first $4$ low-lying states are shown because, when \SAVQD\;was applied to compute
        the first $5$ low-lying states, the ansatz circuits converged to the $1$-st, $2$-nd, $3$-rd,
        $4$-th, and $6$-th states, thereby missing the $5$-th state. The convergence curve of the
        $5$-th state is nearly identical to that of the $4$-th state.
    } 
    \label{fig:lih_results}
\end{figure}

We now present the results for the low-lying $4$ and $5$ energy levels 
of \lih at a near-equilibrium interatomic distance of $1.595$ $\mathring{A}$.
The starting basis set is cc-pVDZ ($19$ orbitals, i.e., $38$ spin-orbitals),
and an active space of $8$ optimized spin-orbitals is used. The
results are shown in \cref{fig:lih_results}.
The overlap terms $\abs{\bra{\Psi_j}U(\bu^{(j)\top}
\bu^{(k)})\ket{\Psi_k}}^2$ are less than $10^{-8}$ at the end of the
optimization. We present the energy convergence curves 
 and terminate the optimization once the convergence
criterion is satisfied. Specifically, convergence is defined as the energy
difference between two consecutive two-step iterations falling below
$2 \times 10^{-5}$ Hartree. Again, we include the state-averaged result
as a straight reference line for comparison.

The weighted energy sum is shown in \cref{tab:lih_total_energy_4}.
In this table, the weight is $4,3,2,1$ for the state $1,2,3,4$.
The \SOVQD\;still gives a better result than the \SAVQD\;in this
case.

\begin{table}[ht]
    \centering
    \begin{tabular}{cccccc}
        \toprule
        State & Method & \textbf{HF} & \SAVQD & \SOVQD & \textbf{FCI} \\
        \midrule
        1 & $E$ (a.u.) & $-8.979$ & $-8.985$ & $-9.000$ & $-9.010$ \\
        2 & $E$ (a.u.) & $-8.834$ & $-8.893$ & $-8.895$ & $-8.896$ \\
        3 & $E$ (a.u.) & $-8.820$ & $-8.873$ & $-8.879$ & $-8.882$ \\
        4 & $E$ (a.u.) & $-8.786$ & $-8.794$ & $-8.856$ & $-8.858$ \\
        \midrule
        Weighted sum & $E$ (a.u.) & $-88.845$ & $-89.160$ & $-89.301$ & $-89.352$ \\
        \bottomrule
    \end{tabular}
    \caption{
        Energies of the first $4$ low-lying states of the \lih molecule 
        in the cc-pVDZ basis set, calculated using \SOVQD\;and \SAVQD. 
        Here the values used for the \SAVQD\;are the first $4$ low-lying states.
        The last row shows the weighted energy sum. Hartree--Fock and FCI values
        are also provided for reference.
    }
    \label{tab:lih_total_energy_4}
\end{table}

In \cref{fig:lih_results}, we present only the first $4$ states. The reason is that
the \SAVQD\;failed to capture the $5$-th state and in fact solved the $6$-th state.
We compute the wavefunction of the $5$-th state by \SAVQD\;and 
compare it with the wavefunction of the $5$-th state by cc-pVDZ FCI. The
overlap is almost $0$ with the $5$-th state by FCI and almost $1$ with the
$6$-th state by FCI. 

The problem arises from relying on a single, insufficiently large set of
orbitals for all states, which prevents an accurate description of states
with significant components on very different orbitals separated by a small
energy gap. In fact, in \lih example, the same partial unitary for the
$\alpha$-spin orbitals and the $\beta$-spin orbitals in \SAVQD\;are initialized to be
\begin{equation*}
    \begin{pmatrix}
        1 & 0 & 0 & 0\\
        0 & 1 & 0 & 0\\
        0 & 0 & 1 & 0\\
        0 & 0 & 0 & 1\\
        0 & 0 & 0 & 0\\
        \vdots & \vdots & \vdots & \vdots\\
        0 & 0 & 0 & 0\\
    \end{pmatrix}.
\end{equation*}
This is appropriate for the first $4$ low-lying states, which have large components
on the first $4$ orbitals. But the $5$-th state has large components on the
$5$-th orbital, i.e., it is very close to the state in \cref{eq:lih_5th_state}.
\begin{equation}\label{eq:lih_5th_state}
        \frac{1}{\sqrt 2} \ket{\underset{19 \,\alpha }{\underbrace{1 1 0 0 \cdots 0}}
        \underset{19\,\beta}{\underbrace{1 0 0 0 1 \cdots 0}}}
        - \frac{1}{\sqrt 2}  \ket{\underset{19\,\alpha}{\underbrace{1 0 0 0 1 \cdots 0}}
        \underset{19\,\beta}{\underbrace{1 1 0 0 \cdots 0}}}.
\end{equation}
At the same time, the $5$-th state energy is the same as the
$4$-th state energy, i.e., $4$-th and $5$-th states are degenerate,
and the $6$-state energy is very close to the $5$-th state energy.
What makes things worse is that the $6$-state has large components
on the first $4$ orbitals. If the state-averaged orbital optimization try to
approximate the $5$-th state, it will away from the first $4$ orbitals, and lose the
accuracy of the first $4$ low-lying states. Thus the state-averaged orbital optimization
algorithm will try to approximate the $6$-th state instead of the $5$-th state.
This problem is more likely to happen when the energy gap between
different states is small.

But for the state-specific orbital optimization algorithm, we can
use different orbitals for different states. The partial unitary matrix
$\bu^{(k)}$ for the $k = 1,2,3,4$ state are initialized to be
\begin{equation*}
    \begin{pmatrix}
        1 & 0 & 0 & 0\\
        0 & 1 & 0 & 0\\
        0 & 0 & 1 & 0\\
        0 & 0 & 0 & 1\\
        0 & 0 & 0 & 0\\
        \vdots & \vdots & \vdots & \vdots\\
        0 & 0 & 0 & 0\\
    \end{pmatrix}
\end{equation*}
and the partial unitary matrix $\bu^{(5)}$ for the $5$-th state
is initialized to be
\begin{equation*}
    \begin{pmatrix}
        1 & 0 & 0 & 0\\
        0 & 1 & 0 & 0\\
        0 & 0 & 1 & 0\\
        0 & 0 & 0 & 0\\
        0 & 0 & 0 & 1\\
        \vdots & \vdots & \vdots & \vdots\\
        0 & 0 & 0 & 0\\
    \end{pmatrix}.
\end{equation*}
In this way, the \SOVQD\;successfully solved the $5$-th state
with the same energy as the $4$-th state.

From this point, we have also tried to use randomly initialized
partial unitary matrix $\bu^{(k)}$ for the $k$-th state, but the
results are worse than the current initialization. This might due to
the fact that this optimization problem is highly nonlinear and
the optimization landscape is very complicated. The randomly initialized
partial unitary matrix $\bu^{(k)}$ may lead to a bad local minimum.

\section{Summary}
\label{sec:summary}
In this paper, we introduced a state-specific orbital optimization scheme
for excited-state calculations on quantum computers. This approach generalizes
the state-averaged orbital optimization scheme by allowing the use of tailored
orbitals for each state, thereby improving both accuracy and flexibility. We
derived the gradient of the overlap term between states generated by different
orbitals and demonstrated how gradient-based optimization methods can be employed
to optimize the orbitals. The scheme was implemented within the Variational
Quantum Deflation (VQD) algorithm, and numerical results on molecules such
as \hhhh and \lih showed that the state-specific orbital optimization scheme
achieves higher accuracy than the state-averaged approach. These results highlight
the potential of state-specific orbital optimization to enhance the performance
of quantum algorithms for electronic structure problems.

\begin{figure}[htb]
    \centering
    \includegraphics{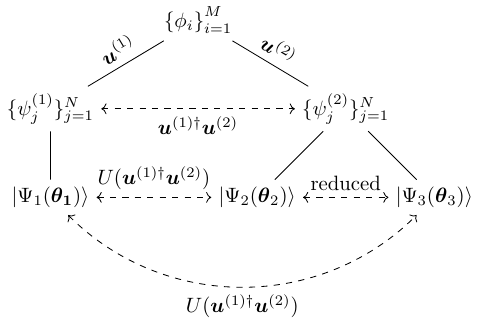}
    \caption{
        The sketch of the state-specific and state-averaged orbital optimization.
        Here the $\ket{\Psi_1(\bthe_1)}$ is the $1$-st state, which uses the 
        orbitals $\{\psi_j^{(1)}\}_{j=1}^N$. $\ket{\Psi_2(\bthe_2)}$ and
        $\ket{\Psi_3(\bthe_3)}$ are the $2$-nd and $3$-rd states, which use
        the same orbitals $\{\psi_j^{(2)}\}_{j=1}^N$. The overlap between
        $\ket{\Psi_2(\bthe_2)}$ and $\ket{\Psi_3(\bthe_3)}$ can be calculated
        straightforwardly without the need of the external circuit like
        $U(\bu^{(1)\dag} \bu^{(2)})$.
        }
    \label{fig:ss_sa_tree}
\end{figure}

As future work, one promising direction is to combine state-averaged and
state-specific strategies. In particular, some states may share a common set of
orbitals while others use individually optimized orbitals. This hybrid strategy
is particularly useful when certain states lie in the same subspace and can be
represented by the same orbitals, such as degenerate states. Beyond this, the
same idea can also be extended to more practical settings, for example enabling
efficient frozen-core calculations where a subset of orbitals is shared or fixed
across multiple states while the remaining orbitals are optimized flexibly. The
sketch of this idea is illustrated in \cref{fig:ss_sa_tree}.

\section*{Acknowledgments}

G.Z. and Y.L. are supported in part
by the National Natural Science Foundation of China (12271109) and the Shanghai Pilot Program for Basic Research-Fudan University 21TQ1400100 (22TQ017). The work of J.B. and J.L. are supported by the US National Science Foundation under award DMS-2309378.

\appendix

\section{State-Averaged Orbital Optimization}
\label{app:state_average}

In this section we will give a short review of the state-averaged orbital
optimization scheme proposed in \cite{bierman2024}.
This method starts from a basis set $\{\phi_i\}_{i=1}^M$ and the rotated orbitals
$\{\psi_j\}_{j=1}^N$ defined with an $M$ times $N$ partial unitary matrix $\bu$ as
\begin{equation*}
    \psi_j = \sum_{i=1}^M  \phi_i \bu_{ij}, \quad \bu^\top \bu = I_N,
\end{equation*}
where $\bu_{ij}$ is the $(i,j)$-th element of the matrix $\bu$.

The problem can be split into two parts: find the
optimal orbitals $\bu$ and find the optimal parameters in the ansatz circuits.
When the orbitals are fixed, the parameters in the ansatz circuits can be
solved by any excited-state solver, such as qOMM, VQD, MCVQE, SSVQE, etc.
When the parameters in the ansatz circuits are fixed, the orbitals can be
optimized by minimizing the weighted average energy of the states defined as
\begin{equation}
    \begin{split}
        F(\bu) = &  \sum_{\alpha=1}^K w_\alpha\big[
        \sum_{p,q=1}^N\sum_{i,j=1}^M h_{ij}\bu_{ip}\bu_{jq} 
        \mathcal{R}^p_{q}(\Psi_\alpha) \\
        &+ \frac{1}{2}\sum_{p,q,r,s=1}^N\sum_{i,j,k,l=1}^M
        v_{pqrs}\bu_{ip}\bu_{jq}\bu_{kr}\bu_{ls}
        \mathcal{R}^{pq}_{rs}(\Psi_\alpha)\big]\\
        = & \sum_{\alpha=1}^K w_\alpha \textbf{E}_{\text{SS}}^\alpha(\bthe_\alpha,\bu),
    \end{split}
\end{equation}
where
$\bthe_\alpha$ are the parameters in the ansatz circuit for the $\alpha$-th state,
$w_\alpha > 0$ are the weights for the $\alpha$-th state.

As demonstrated in \cite{li2020,bierman2023,bierman2024}, constrained
projected gradient descent is an effective optimization strategy for the degree-$4$ polynomial
objective $F(\bu)$ involving the partial unitary matrix $\bu$,
which has a parameter update step as
\begin{equation}
    \bu \gets \text{Orth}(\bu - \eta \nabla_{\bu} F(\bu)),
\end{equation}
where $\eta$ is the step size and $\nabla_{\bu} F(\bu)$ is the gradient of
the function $F(\bu)$ with respect to the partial unitary matrix $\bu$.
The orthogonalization can be
chosen to be the Q matrix in QR decomposition, or the product of the left and right
singular vectors in singular value decomposition~(SVD).
In this paper, we use SVD for orthogonalization, i.e., 
\begin{equation}
    \text{Orth}(A) = U V^\top, \text{where } A = U \Sigma V^\top.
\end{equation}

A key point of this method is that the optimization of the orbitals can be
done on classical computers after all of the $1$-RDMs and $2$-RDMs are calculated
by quantum circuits. This is different with the state-specific orbital optimization,
where quantum circuits are needed to calculate the gradients. We will discuss it
in next section.

In summary, the state-averaged orbital optimization can be formulated as
the following pseudo-code \cref{alg:state_averaged_orbital_optimization}.
\begin{algorithm}[h]
    \caption{State-Averaged Orbital Optimization}
    \label{alg:state_averaged_orbital_optimization}
    \KwIn{$\mathcal{H}$, $\bu_{\text{init}}$}
    \KwOut{$\{\Psi_\alpha\}_{\alpha=1}^K$, $\bu_{\text{opt}}$}
    Initialize $\bu = \bu_{\text{init}}$\;
    \While{not converged}{
        Freeze $\bu$, call a quantum eigensolver to obtian
        $\{\Psi_\alpha\}_{\alpha=1}^K$\;
        Calculate the $\mathcal{R}^p_{q}(\Psi_\alpha)$ and
        $\mathcal{R}^{pq}_{rs}(\Psi_\alpha)$ for all $\alpha = 1, 2, \dots, K$
        , $p, q, r, s = 1, 2, \dots, N$ by quantum circuits\;
        \While{not converged}{
            $\bu \gets \text{Orth}(\bu - \eta \nabla_{\bu} F(\bu))$,
        }
    }
    Return $\{\Psi_\alpha\}_{\alpha=1}^K$, $\bu$\;
\end{algorithm}

The state-averaged algorithm
requires careful selection of the number of active orbitals $N$, as an
inappropriate choice may result in the loss of certain excited states.
A numerical example illustrating this issue is also provided
in \cref{sec:lih}.

\section{Gradient of Overlap}
\label{app:gradient_of_overlap}

In this section, we will derive the gradient of the linear operator
$U(\bu)$ with respect to $\bu$ and then use it to calculate the gradient
of the overlap between two many-body wave functions in different basis sets.
We will use the exterior algebra to represent the many-body wave functions
and the annihilators and creators of the many-body wave functions, which
are also used in \cite{zhu2025}.

\subsection{Notations}

Given a vector space $\mathcal{V}$ and a basis set $\{\psi_i\}_{i=1}^n$,
then the $i$-th annihilator $\ba_i = \ba(\psi_i)$ and
creator $\ba_i^\dag = \ba^\dag(\psi_i)$ are defined as follows:
\begin{align*}
    &\ba_i^\dag(w) = \psi_i \wedge w, \quad \text{for any } w \in \wedge V, \\
    &\ba_i(\psi_{i_1} \wedge \psi_{i_2} \wedge \cdots \wedge \psi_{i_k}) \\
    = & 
    \begin{cases}
        &{(-1)}^{\textbf{n}(i)} \cdot \psi_{i_1} \wedge \psi_{i_2} \wedge
        \cdots \wedge {\hat{\psi}}_i \wedge \cdots \wedge \psi_{i_k}, \\
        & \text{if\;} i \in \{i_1, i_2, \dots, i_k\}, \\
        &0, \\
        & \text{if\;} i \notin \{i_1, i_2, \dots, i_k\}.
    \end{cases}
\end{align*}

Here, $\wedge \mathcal{V}$ is the exterior algebra of $\mathcal{V}$,
which is the space of antisymmetric tensors, and
$\psi_{i_1} \wedge \psi_{i_2} \wedge \cdots \wedge \psi_{i_k}$ is a
$k$-form in $\wedge \mathcal{V}$. Here we use the convention that the
lower index of the wedge product is grown from left to right, i.e.,
$1 \le i_1 < i_2 < \cdots < i_k \le n$. The notation ${\hat{\psi}}_i$ means 
that the term $\psi_i$ is omitted from the wedge product. $\textbf{n}(i)$
is the number of terms in the wedge product which index is less than $i$,
i.e.,
\begin{equation*}
    \textbf{n}(i) = \sum_{i_s < i} 1,
\end{equation*}
where $i_s$ are the indices of the terms in the wedge product.

This definition of annihilators and creators is consistent with the
definition of annihilators and creators in quantum mechanics. You can think
the $k$-form $\psi_{i_1} \wedge \psi_{i_2} \wedge \cdots \wedge \psi_{i_k}$
as a many-body wave function, where the $k$ particles are in the states
$\psi_{i_1}, \psi_{i_2}, \dots, \psi_{i_k}$. The annihilator $\ba_i$ removes the
particle in the state $\psi_i$ from the many-body wave function, and the
creator $\ba_i^\dag$ adds a particle in the state $\psi_i$ to the
many-body wave function.

For a linear operator $\bu$ on $\mathcal{V}$ with matrix representation $\bu_{ji}$
defined as follows:
\begin{equation*}
    \bu\psi_i = \sum_j \psi_j \bu_{ji}, \quad \bu_{ji} = \bra{\psi_j}\bu\ket{\psi_i},
\end{equation*}
we will use the notation $U(\bu)$ to denote the linear operator acting on
$\wedge \mathcal{V}$ extended from the linear operator $\bu$ as follows:
\begin{equation}\label{eq:U_bu}
    \begin{split}
        &U(\bu)(\psi_{i_1} \wedge \psi_{i_2} \wedge \cdots \wedge \psi_{i_k}) \\
        = & (\bu\psi_{i_1}) \wedge (\bu\psi_{i_2}) \wedge \cdots \wedge
        (\bu\psi_{i_k}) \\
        = & \sum_{j_1, j_2, \dots, j_k} \bu_{j_1 i_1} \bu_{j_2 i_2} \cdots
        \bu_{j_k i_k}
        \cdot \psi_{j_1} \wedge \psi_{j_2} \wedge \cdots \wedge \psi_{j_k}.
    \end{split}
\end{equation}

These notations will be used in the following section to derive the
gradient of the overlap between two many-body wave functions.

\subsection{Gradient Expression}

Now for any two linear operators $\bu$ and $\bv$ on $\mathcal{V}$, we can
calculate the difference of the linear operator $U(\bu + \bv)$ and
$U(\bu)$ acting on a $k$-form $\psi_{i_1} \wedge \psi_{i_2} \wedge
\cdots \wedge \psi_{i_k}$ by expanding the definition \cref{eq:U_bu}:
\begin{equation}\label{eq:U_bu_plus_v_minus_U_bu}
    \begin{split}
        &(U(\bu + \bv) - U(\bu)) \psi_{i_1} \wedge \psi_{i_2} \wedge \cdots \wedge \psi_{i_k} \\
        = &\big(\bv\psi_{i_1} \wedge \bu\psi_{i_2} \wedge \cdots \wedge \bu\psi_{i_k}\big) \\
        &+ \big(\bu\psi_{i_1} \wedge \bv\psi_{i_2} \wedge \cdots \wedge \bu\psi_{i_k}\big) \\
        &+ \cdots \\
        &+ \big(\bu\psi_{i_1} \wedge \bu\psi_{i_2} \wedge \cdots \wedge \bv\psi_{i_k}\big) \\
        &+ o(\norm{\bv}).
    \end{split}
\end{equation}

Now we consider the action of the linear operator
\begin{equation*}
    \sum_{ji} \bv_{ji} \ba_j^\dag U(\bu) \ba_i
\end{equation*}
on the $k$-form $\psi_{i_1} \wedge \psi_{i_2} \wedge \cdots \wedge \psi_{i_k}$,
where $\bv_{ji} = \bra{\psi_j} \bv \ket{\psi_i}$, $\ba_j = \ba(\psi_j)$, and
$\ba_i^\dag = \ba^\dag(\psi_i)$. The action of this linear operator can be
calculated by cases:
\begin{enumerate}
    \item If $i \notin \{i_1, i_2, \cdots, i_k\}$, then
            \begin{equation*}
                [\sum_{j}v_{ji}\ba_j^\dag U(\bu)\ba_i]
                ( \psi_{i_1}\wedge\psi_{i_2}\wedge\cdots\wedge\psi_{i_k}) = 0.
            \end{equation*}
    \item If $i \in \{i_1, i_2, \cdots, i_k\}$, sthen we have
        \begin{align*}
            &[\sum_{j}\bv_{ji}\ba_j^\dag U(\bu)\ba_i]
            ( \psi_{i_1}\wedge\psi_{i_2}\wedge\cdots\wedge\psi_{i_k}) \\
            = & {(-1)}^{\textbf{n}(i)}\sum_{j}\bv_{ji}\ba_j^\dag U(\bu)(\psi_{i_1}
            \wedge \psi_{i_2}\wedge\cdots\wedge{\hat{\psi}}_i\wedge\cdots\wedge \psi_{i_k}) \\
            = & {(-1)}^{\textbf{n}(i)}\sum_{j}\bv_{ji}\ba_j^\dag(\bu\psi_{i_1}
            \wedge \bu\psi_{i_2}\wedge\cdots\wedge{\hat{\psi}}_i\wedge\cdots\wedge \bu\psi_{i_k}) \\
            = & {(-1)}^{\textbf{n}(i)}(\sum_{j}\psi_j \bv_{ji})\wedge \bu\psi_{i_1}
            \wedge \bu\psi_{i_2}\wedge\cdots\wedge{\hat{\psi}}_i\wedge\cdots\wedge \bu\psi_{i_k} \\
            = & \bu\psi_{i_1}\wedge \bu\psi_{i_2}\wedge\cdots\wedge \bv\psi_i\wedge 
            \cdots\wedge \bu\psi_{i_k}.
        \end{align*}
\end{enumerate}

Merge the two cases above, we have
\begin{align*}
    &\left[\sum_{ji} \bv_{ji} \ba_j^\dag U(\bu) \ba_i\right] 
    \left( \psi_{i_1} \wedge \psi_{i_2} \wedge \cdots \wedge \psi_{i_k} \right) \\
    = & \sum_{s=1}^k 
    \left( \bu\psi_{i_1} \wedge \cdots \wedge \bv\psi_{i_s} \wedge \cdots 
    \wedge \bu\psi_{i_k} \right),
\end{align*}
which is the same as the first term in \cref{eq:U_bu_plus_v_minus_U_bu}.
As a result, we have
\begin{align*}
    & (U(\bu+\bv)-U(\bu))\psi_{i_1}\wedge\psi_{i_2}\wedge\cdots\wedge\psi_{i_k}\\
    = & [\sum_{ji}\bv_{ji}\ba_j^\dag U(\bu)\ba_i]( \psi_{i_1}\wedge\psi_{i_2}
    \wedge\cdots\wedge\psi_{i_k}) + o(\norm{\bv}).
\end{align*}

Therefore, by the definition of the derivative of a linear operator, we have
\begin{equation}\label{eq:gradient_of_U_bu_op}
    \dd (U(\bu)) = \sum_{ij}(\dd \bu)_{ij}\ba_i^\dag U(\bu)\ba_j.
\end{equation}
Or in matrix form, we have
\begin{equation}\label{eq:gradient_of_U_bu_mat}
    \frac{\partial}{\partial \bu_{ij}} U(\bu) = \sum_{ij} \ba_i^\dag U(\bu) \ba_j,
\end{equation}
Here $\dd \bu$ is a the derivative of $\bu$ and
$(\dd \bu)_{ij} = \bra{\psi_i}\dd \bu\ket{\psi_j}$, $\ba_i = \ba(\psi_i)$,
$\ba_i^\dag = \ba^\dag(\psi_i)$.

Now for the overlap between two many-body wave functions $\ket{\Psi_1}$ and
$\ket{\Psi_2}$ with real basis rotations $\bu^{(1)}$ and $\bu^{(2)}$, \cite{zhu2025}
showed that the overlap can be expressed as
\begin{equation*}
    \abs{\bra{\Psi_1}U(\bu^{(1)\top}\bu^{(2)})\ket{\Psi_2}}^2.
\end{equation*}
Take derivative of the overlap with respect to $\bu^{(2)}_{pq}$ and use
\cref{eq:gradient_of_U_bu_mat} with chain rule, we have
\begin{equation*}
    \begin{split}
        &\frac{\partial}{\partial \bu^{(2)}_{pq}}|\bra{\Psi_1}
        U(\bu^{(1)\top}\bu^{(2)})\ket{\Psi_2}|^2 \\
        = &\frac{\partial}{\partial \bu^{(2)}_{pq}}\big( \bra{\Psi_1}
        U(\bu^{(1)\top}\bu^{(2)})\ket{\Psi_2}\bra{\Psi_2}
        U(\bu^{(2)\top}\bu^{(1)})\ket{\Psi_1}\big) \\
        = & \bra{\Psi_2}U(\bu^{(2)\top}\bu^{(1)}) \ket{\Psi_1} \\
          &\cdot \frac{\partial}{\partial \bu^{(2)}_{pq}}
        \bra{\Psi_1}U(\bu^{(1)\top}\bu^{(2)})\ket{\Psi_2} + \text{c.c.}\\
        = & \bra{\Psi_2}U(\bu^{(2)\top}\bu^{(1)}) \ket{\Psi_1} \\
          &\cdot \sum_{ij}
          \frac{\partial (\bu^{(1)\top}\bu^{(2)})_{ij}}{\partial \bu^{(2)}_{pq}}\bra{\Psi_1}
          \ba_i^\dag U(\bu^{(1)\top}\bu^{(2)})\ba_j\ket{\Psi_2} + \text{c.c.}\\
        = & \bra{\Psi_2}U(\bu^{(2)\top}\bu^{(1)}) \ket{\Psi_1} \\
          &\cdot \sum_{ijk}
          \frac{\partial (\bu^{(1)}_{ki}\bu^{(2)}_{kj})}{\partial \bu^{(2)}_{pq}}\bra{\Psi_1}
          \ba_i^\dag U(\bu^{(1)\top}\bu^{(2)})\ba_j\ket{\Psi_2} + \text{c.c.}\\
        = & \bra{\Psi_2}U(\bu^{(2)\top}\bu^{(1)}) \ket{\Psi_1} \\
          &\cdot \sum_{ijk}
          \left(\bu^{(1)}_{ki}\delta_{kp}\delta_{jq}\right)
          \bra{\Psi_1} \ba_i^\dag U(\bu^{(1)\top}\bu^{(2)})\ba_j\ket{\Psi_2} + \text{c.c.}\\
        = & \bra{\Psi_2}U(\bu^{(2)\top}\bu^{(1)}) \ket{\Psi_1} \\
          &\cdot \sum_{i}
          \bu^{(1)}_{pi}
          \bra{\Psi_1} \ba_i^\dag U(\bu^{(1)\top}\bu^{(2)})\ba_q\ket{\Psi_2} + \text{c.c.}\\
    \end{split}
\end{equation*}

Thus, we have the gradient of the overlap between two many-body wave functions
$\ket{\Psi_1}$ and $\ket{\Psi_2}$ with real basis rotations $\bu^1$ and $\bu^2$ as
\begin{equation*}
    \begin{split}
        &\frac{\partial}{\partial \bu^{(2)}_{pq}}|\bra{\Psi_1}
        U(\bu^{(1)\top}\bu^{(2)})\ket{\Psi_2}|^2 \\
        = & \bra{\Psi_2}U(\bu^{(2)\top}\bu^{(1)}) \ket{\Psi_1} \\
          &\cdot \sum_{i}
          \bu^{(1)}_{pi}
          \bra{\Psi_1} \ba_i^\dag U(\bu^{(1)\top}\bu^{(2)})\ba_q\ket{\Psi_2} + \text{c.c.}\\
    \end{split}
\end{equation*}
which the same as \cref{eq:gradient_of_overlap} in the main text.

\clearpage
\bibliography{reference}

\begin{thebibliography}{32}%
\makeatletter
\providecommand \@ifxundefined [1]{%
 \@ifx{#1\undefined}
}%
\providecommand \@ifnum [1]{%
 \ifnum #1\expandafter \@firstoftwo
 \else \expandafter \@secondoftwo
 \fi
}%
\providecommand \@ifx [1]{%
 \ifx #1\expandafter \@firstoftwo
 \else \expandafter \@secondoftwo
 \fi
}%
\providecommand \natexlab [1]{#1}%
\providecommand \enquote  [1]{``#1''}%
\providecommand \bibnamefont  [1]{#1}%
\providecommand \bibfnamefont [1]{#1}%
\providecommand \citenamefont [1]{#1}%
\providecommand \href@noop [0]{\@secondoftwo}%
\providecommand \href [0]{\begingroup \@sanitize@url \@href}%
\providecommand \@href[1]{\@@startlink{#1}\@@href}%
\providecommand \@@href[1]{\endgroup#1\@@endlink}%
\providecommand \@sanitize@url [0]{\catcode `\\12\catcode `\$12\catcode `\&12\catcode `\#12\catcode `\^12\catcode `\_12\catcode `\%12\relax}%
\providecommand \@@startlink[1]{}%
\providecommand \@@endlink[0]{}%
\providecommand \url  [0]{\begingroup\@sanitize@url \@url }%
\providecommand \@url [1]{\endgroup\@href {#1}{\urlprefix }}%
\providecommand \urlprefix  [0]{URL }%
\providecommand \Eprint [0]{\href }%
\providecommand \doibase [0]{https://doi.org/}%
\providecommand \selectlanguage [0]{\@gobble}%
\providecommand \bibinfo  [0]{\@secondoftwo}%
\providecommand \bibfield  [0]{\@secondoftwo}%
\providecommand \translation [1]{[#1]}%
\providecommand \BibitemOpen [0]{}%
\providecommand \bibitemStop [0]{}%
\providecommand \bibitemNoStop [0]{.\EOS\space}%
\providecommand \EOS [0]{\spacefactor3000\relax}%
\providecommand \BibitemShut  [1]{\csname bibitem#1\endcsname}%
\let\auto@bib@innerbib\@empty
\bibitem [{\citenamefont {Helgaker}\ \emph {et~al.}(2013)\citenamefont {Helgaker}, \citenamefont {Jorgensen},\ and\ \citenamefont {Olsen}}]{helgaker2013molecular}%
  \BibitemOpen
  \bibfield  {author} {\bibinfo {author} {\bibfnamefont {T.}~\bibnamefont {Helgaker}}, \bibinfo {author} {\bibfnamefont {P.}~\bibnamefont {Jorgensen}},\ and\ \bibinfo {author} {\bibfnamefont {J.}~\bibnamefont {Olsen}},\ }\href@noop {} {\emph {\bibinfo {title} {Molecular electronic-structure theory}}}\ (\bibinfo  {publisher} {John Wiley \& Sons},\ \bibinfo {year} {2013})\BibitemShut {NoStop}%
\bibitem [{\citenamefont {Szabo}\ and\ \citenamefont {Ostlund}(1996)}]{szabo1996modern}%
  \BibitemOpen
  \bibfield  {author} {\bibinfo {author} {\bibfnamefont {A.}~\bibnamefont {Szabo}}\ and\ \bibinfo {author} {\bibfnamefont {N.~S.}\ \bibnamefont {Ostlund}},\ }\href@noop {} {\emph {\bibinfo {title} {Modern quantum chemistry: introduction to advanced electronic structure theory}}}\ (\bibinfo  {publisher} {Courier Corporation},\ \bibinfo {year} {1996})\BibitemShut {NoStop}%
\bibitem [{\citenamefont {Peruzzo}\ \emph {et~al.}(2014)\citenamefont {Peruzzo}, \citenamefont {McClean}, \citenamefont {Shadbolt}, \citenamefont {Yung}, \citenamefont {Zhou}, \citenamefont {Love}, \citenamefont {Aspuru-Guzik},\ and\ \citenamefont {O’Brien}}]{peruzzo2014variational}%
  \BibitemOpen
  \bibfield  {author} {\bibinfo {author} {\bibfnamefont {A.}~\bibnamefont {Peruzzo}}, \bibinfo {author} {\bibfnamefont {J.}~\bibnamefont {McClean}}, \bibinfo {author} {\bibfnamefont {P.}~\bibnamefont {Shadbolt}}, \bibinfo {author} {\bibfnamefont {M.-H.}\ \bibnamefont {Yung}}, \bibinfo {author} {\bibfnamefont {X.-Q.}\ \bibnamefont {Zhou}}, \bibinfo {author} {\bibfnamefont {P.~J.}\ \bibnamefont {Love}}, \bibinfo {author} {\bibfnamefont {A.}~\bibnamefont {Aspuru-Guzik}},\ and\ \bibinfo {author} {\bibfnamefont {J.~L.}\ \bibnamefont {O’Brien}},\ }\bibfield  {title} {\bibinfo {title} {A variational eigenvalue solver on a photonic quantum processor},\ }\href {https://doi.org/10.1038/ncomms5213} {\bibfield  {journal} {\bibinfo  {journal} {Nature Communications}\ }\textbf {\bibinfo {volume} {5}},\ \bibinfo {pages} {4213} (\bibinfo {year} {2014})}\BibitemShut {NoStop}%
\bibitem [{\citenamefont {Tilly}\ \emph {et~al.}(2022)\citenamefont {Tilly}, \citenamefont {Chen}, \citenamefont {Cao}, \citenamefont {Picozzi}, \citenamefont {Setia}, \citenamefont {Li}, \citenamefont {Grant}, \citenamefont {Wossnig}, \citenamefont {Rungger}, \citenamefont {Booth},\ and\ \citenamefont {Tennyson}}]{tilly2022variational}%
  \BibitemOpen
  \bibfield  {author} {\bibinfo {author} {\bibfnamefont {J.}~\bibnamefont {Tilly}}, \bibinfo {author} {\bibfnamefont {H.}~\bibnamefont {Chen}}, \bibinfo {author} {\bibfnamefont {S.}~\bibnamefont {Cao}}, \bibinfo {author} {\bibfnamefont {D.}~\bibnamefont {Picozzi}}, \bibinfo {author} {\bibfnamefont {K.}~\bibnamefont {Setia}}, \bibinfo {author} {\bibfnamefont {Y.}~\bibnamefont {Li}}, \bibinfo {author} {\bibfnamefont {E.}~\bibnamefont {Grant}}, \bibinfo {author} {\bibfnamefont {L.}~\bibnamefont {Wossnig}}, \bibinfo {author} {\bibfnamefont {I.}~\bibnamefont {Rungger}}, \bibinfo {author} {\bibfnamefont {G.~H.}\ \bibnamefont {Booth}},\ and\ \bibinfo {author} {\bibfnamefont {J.}~\bibnamefont {Tennyson}},\ }\bibfield  {title} {\bibinfo {title} {The {Variational} {Quantum} {Eigensolver}: {A} review of methods and best practices},\ }\href {https://doi.org/https://doi.org/10.1016/j.physrep.2022.08.003} {\bibfield  {journal} {\bibinfo  {journal} {Physics Reports}\ }\textbf {\bibinfo {volume} {986}},\ \bibinfo {pages} {1}
  (\bibinfo {year} {2022})}\BibitemShut {NoStop}%
\bibitem [{\citenamefont {Nielsen}\ and\ \citenamefont {Chuang}(2010)}]{nielsen2010quantum}%
  \BibitemOpen
  \bibfield  {author} {\bibinfo {author} {\bibfnamefont {M.~A.}\ \bibnamefont {Nielsen}}\ and\ \bibinfo {author} {\bibfnamefont {I.~L.}\ \bibnamefont {Chuang}},\ }\href@noop {} {\emph {\bibinfo {title} {Quantum computation and quantum information}}}\ (\bibinfo  {publisher} {Cambridge university press},\ \bibinfo {year} {2010})\BibitemShut {NoStop}%
\bibitem [{\citenamefont {Kitaev}(1995)}]{kitaev1995qpe}%
  \BibitemOpen
  \bibfield  {author} {\bibinfo {author} {\bibfnamefont {A.~Y.}\ \bibnamefont {Kitaev}},\ }\bibfield  {title} {\bibinfo {title} {Quantum measurements and the abelian stabilizer problem},\ }\href@noop {} {\bibfield  {journal} {\bibinfo  {journal} {arXiv preprint quant-ph/9511026}\ } (\bibinfo {year} {1995})}\BibitemShut {NoStop}%
\bibitem [{\citenamefont {Kühn}\ \emph {et~al.}(2019)\citenamefont {Kühn}, \citenamefont {Zanker}, \citenamefont {Deglmann}, \citenamefont {Marthaler},\ and\ \citenamefont {Weiß}}]{kühn2019accuracy}%
  \BibitemOpen
  \bibfield  {author} {\bibinfo {author} {\bibfnamefont {M.}~\bibnamefont {Kühn}}, \bibinfo {author} {\bibfnamefont {S.}~\bibnamefont {Zanker}}, \bibinfo {author} {\bibfnamefont {P.}~\bibnamefont {Deglmann}}, \bibinfo {author} {\bibfnamefont {M.}~\bibnamefont {Marthaler}},\ and\ \bibinfo {author} {\bibfnamefont {H.}~\bibnamefont {Weiß}},\ }\bibfield  {title} {\bibinfo {title} {Accuracy and resource estimations for quantum chemistry on a near-term quantum computer},\ }\href {https://doi.org/10.1021/acs.jctc.9b00236} {\bibfield  {journal} {\bibinfo  {journal} {Journal of Chemical Theory and Computation}\ }\textbf {\bibinfo {volume} {15}},\ \bibinfo {pages} {4764–4780} (\bibinfo {year} {2019})}\BibitemShut {NoStop}%
\bibitem [{\citenamefont {Elfving}\ \emph {et~al.}(2020)\citenamefont {Elfving}, \citenamefont {Broer}, \citenamefont {Webber}, \citenamefont {Gavartin}, \citenamefont {Halls}, \citenamefont {Lorton},\ and\ \citenamefont {Bochevarov}}]{elfving2020will}%
  \BibitemOpen
  \bibfield  {author} {\bibinfo {author} {\bibfnamefont {V.~E.}\ \bibnamefont {Elfving}}, \bibinfo {author} {\bibfnamefont {B.~W.}\ \bibnamefont {Broer}}, \bibinfo {author} {\bibfnamefont {M.}~\bibnamefont {Webber}}, \bibinfo {author} {\bibfnamefont {J.}~\bibnamefont {Gavartin}}, \bibinfo {author} {\bibfnamefont {M.~D.}\ \bibnamefont {Halls}}, \bibinfo {author} {\bibfnamefont {K.~P.}\ \bibnamefont {Lorton}},\ and\ \bibinfo {author} {\bibfnamefont {A.}~\bibnamefont {Bochevarov}},\ }\bibfield  {title} {\bibinfo {title} {How will quantum computers provide an industrially relevant computational advantage in quantum chemistry?},\ }\href@noop {} {\bibfield  {journal} {\bibinfo  {journal} {arXiv preprint arXiv:2009.12472}\ } (\bibinfo {year} {2020})}\BibitemShut {NoStop}%
\bibitem [{\citenamefont {Gonthier}\ \emph {et~al.}(2022)\citenamefont {Gonthier}, \citenamefont {Radin}, \citenamefont {Buda}, \citenamefont {Doskocil}, \citenamefont {Abuan},\ and\ \citenamefont {Romero}}]{gonthier2022measurements}%
  \BibitemOpen
  \bibfield  {author} {\bibinfo {author} {\bibfnamefont {J.~F.}\ \bibnamefont {Gonthier}}, \bibinfo {author} {\bibfnamefont {M.~D.}\ \bibnamefont {Radin}}, \bibinfo {author} {\bibfnamefont {C.}~\bibnamefont {Buda}}, \bibinfo {author} {\bibfnamefont {E.~J.}\ \bibnamefont {Doskocil}}, \bibinfo {author} {\bibfnamefont {C.~M.}\ \bibnamefont {Abuan}},\ and\ \bibinfo {author} {\bibfnamefont {J.}~\bibnamefont {Romero}},\ }\bibfield  {title} {\bibinfo {title} {Measurements as a roadblock to near-term practical quantum advantage in chemistry: Resource analysis},\ }\href@noop {} {\bibfield  {journal} {\bibinfo  {journal} {Physical Review Research}\ }\textbf {\bibinfo {volume} {4}},\ \bibinfo {pages} {033154} (\bibinfo {year} {2022})}\BibitemShut {NoStop}%
\bibitem [{\citenamefont {Roos}\ \emph {et~al.}(1980)\citenamefont {Roos}, \citenamefont {Taylor},\ and\ \citenamefont {Sigbahn}}]{roos1980casscf}%
  \BibitemOpen
  \bibfield  {author} {\bibinfo {author} {\bibfnamefont {B.~O.}\ \bibnamefont {Roos}}, \bibinfo {author} {\bibfnamefont {P.~R.}\ \bibnamefont {Taylor}},\ and\ \bibinfo {author} {\bibfnamefont {P.~E.}\ \bibnamefont {Sigbahn}},\ }\bibfield  {title} {\bibinfo {title} {A complete active space scf method (casscf) using a density matrix formulated super-ci approach},\ }\href@noop {} {\bibfield  {journal} {\bibinfo  {journal} {Chemical Physics}\ }\textbf {\bibinfo {volume} {48}},\ \bibinfo {pages} {157} (\bibinfo {year} {1980})}\BibitemShut {NoStop}%
\bibitem [{\citenamefont {Siegbahn}\ \emph {et~al.}(1981)\citenamefont {Siegbahn}, \citenamefont {Alml{\"o}f}, \citenamefont {Heiberg},\ and\ \citenamefont {Roos}}]{siegbahn1981casscf}%
  \BibitemOpen
  \bibfield  {author} {\bibinfo {author} {\bibfnamefont {P.~E.}\ \bibnamefont {Siegbahn}}, \bibinfo {author} {\bibfnamefont {J.}~\bibnamefont {Alml{\"o}f}}, \bibinfo {author} {\bibfnamefont {A.}~\bibnamefont {Heiberg}},\ and\ \bibinfo {author} {\bibfnamefont {B.~O.}\ \bibnamefont {Roos}},\ }\bibfield  {title} {\bibinfo {title} {The complete active space scf (casscf) method in a newton--raphson formulation with application to the hno molecule},\ }\href@noop {} {\bibfield  {journal} {\bibinfo  {journal} {The Journal of Chemical Physics}\ }\textbf {\bibinfo {volume} {74}},\ \bibinfo {pages} {2384} (\bibinfo {year} {1981})}\BibitemShut {NoStop}%
\bibitem [{\citenamefont {Siegbahn}\ \emph {et~al.}(1980)\citenamefont {Siegbahn}, \citenamefont {Heiberg}, \citenamefont {Roos},\ and\ \citenamefont {Levy}}]{siegbahn1980comparison}%
  \BibitemOpen
  \bibfield  {author} {\bibinfo {author} {\bibfnamefont {P.}~\bibnamefont {Siegbahn}}, \bibinfo {author} {\bibfnamefont {A.}~\bibnamefont {Heiberg}}, \bibinfo {author} {\bibfnamefont {B.}~\bibnamefont {Roos}},\ and\ \bibinfo {author} {\bibfnamefont {B.}~\bibnamefont {Levy}},\ }\bibfield  {title} {\bibinfo {title} {A comparison of the super-ci and the newton-raphson scheme in the complete active space scf method},\ }\href@noop {} {\bibfield  {journal} {\bibinfo  {journal} {Physica Scripta}\ }\textbf {\bibinfo {volume} {21}},\ \bibinfo {pages} {323} (\bibinfo {year} {1980})}\BibitemShut {NoStop}%
\bibitem [{\citenamefont {Li}\ and\ \citenamefont {Lu}(2020)}]{li2020}%
  \BibitemOpen
  \bibfield  {author} {\bibinfo {author} {\bibfnamefont {Y.}~\bibnamefont {Li}}\ and\ \bibinfo {author} {\bibfnamefont {J.}~\bibnamefont {Lu}},\ }\bibfield  {title} {\bibinfo {title} {Optimal orbital selection for full configuration interaction (optorbfci): Pursuing the basis set limit under a budget},\ }\href@noop {} {\bibfield  {journal} {\bibinfo  {journal} {Journal of chemical theory and computation}\ }\textbf {\bibinfo {volume} {16}},\ \bibinfo {pages} {6207} (\bibinfo {year} {2020})}\BibitemShut {NoStop}%
\bibitem [{\citenamefont {Yalouz}\ \emph {et~al.}(2021)\citenamefont {Yalouz}, \citenamefont {Senjean}, \citenamefont {Günther}, \citenamefont {Buda}, \citenamefont {O’Brien},\ and\ \citenamefont {Visscher}}]{yalouz2021saoo}%
  \BibitemOpen
  \bibfield  {author} {\bibinfo {author} {\bibfnamefont {S.}~\bibnamefont {Yalouz}}, \bibinfo {author} {\bibfnamefont {B.}~\bibnamefont {Senjean}}, \bibinfo {author} {\bibfnamefont {J.}~\bibnamefont {Günther}}, \bibinfo {author} {\bibfnamefont {F.}~\bibnamefont {Buda}}, \bibinfo {author} {\bibfnamefont {T.~E.}\ \bibnamefont {O’Brien}},\ and\ \bibinfo {author} {\bibfnamefont {L.}~\bibnamefont {Visscher}},\ }\bibfield  {title} {\bibinfo {title} {A state-averaged orbital-optimized hybrid quantum–classical algorithm for a democratic description of ground and excited states},\ }\href {https://doi.org/10.1088/2058-9565/abd334} {\bibfield  {journal} {\bibinfo  {journal} {Quantum Science and Technology}\ }\textbf {\bibinfo {volume} {6}},\ \bibinfo {pages} {024004} (\bibinfo {year} {2021})},\ \bibinfo {note} {publisher: IOP Publishing}\BibitemShut {NoStop}%
\bibitem [{\citenamefont {Omiya}\ \emph {et~al.}(2022)\citenamefont {Omiya}, \citenamefont {Nakagawa}, \citenamefont {Koh}, \citenamefont {Mizukami}, \citenamefont {Gao},\ and\ \citenamefont {Kobayashi}}]{omiya2022saoo}%
  \BibitemOpen
  \bibfield  {author} {\bibinfo {author} {\bibfnamefont {K.}~\bibnamefont {Omiya}}, \bibinfo {author} {\bibfnamefont {Y.~O.}\ \bibnamefont {Nakagawa}}, \bibinfo {author} {\bibfnamefont {S.}~\bibnamefont {Koh}}, \bibinfo {author} {\bibfnamefont {W.}~\bibnamefont {Mizukami}}, \bibinfo {author} {\bibfnamefont {Q.}~\bibnamefont {Gao}},\ and\ \bibinfo {author} {\bibfnamefont {T.}~\bibnamefont {Kobayashi}},\ }\bibfield  {title} {\bibinfo {title} {Analytical {Energy} {Gradient} for {State}-{Averaged} {Orbital}-{Optimized} {Variational} {Quantum} {Eigensolvers} and {Its} {Application} to a {Photochemical} {Reaction}},\ }\href {https://doi.org/10.1021/acs.jctc.1c00877} {\bibfield  {journal} {\bibinfo  {journal} {Journal of Chemical Theory and Computation}\ }\textbf {\bibinfo {volume} {18}},\ \bibinfo {pages} {741} (\bibinfo {year} {2022})},\ \bibinfo {note} {\_eprint: https://doi.org/10.1021/acs.jctc.1c00877}\BibitemShut {NoStop}%
\bibitem [{\citenamefont {Bierman}\ \emph {et~al.}(2023)\citenamefont {Bierman}, \citenamefont {Li},\ and\ \citenamefont {Lu}}]{bierman2023}%
  \BibitemOpen
  \bibfield  {author} {\bibinfo {author} {\bibfnamefont {J.}~\bibnamefont {Bierman}}, \bibinfo {author} {\bibfnamefont {Y.}~\bibnamefont {Li}},\ and\ \bibinfo {author} {\bibfnamefont {J.}~\bibnamefont {Lu}},\ }\bibfield  {title} {\bibinfo {title} {Improving the accuracy of variational quantum eigensolvers with fewer qubits using orbital optimization},\ }\href@noop {} {\bibfield  {journal} {\bibinfo  {journal} {Journal of Chemical Theory and Computation}\ }\textbf {\bibinfo {volume} {19}},\ \bibinfo {pages} {790} (\bibinfo {year} {2023})}\BibitemShut {NoStop}%
\bibitem [{\citenamefont {Bierman}\ \emph {et~al.}(2024)\citenamefont {Bierman}, \citenamefont {Li},\ and\ \citenamefont {Lu}}]{bierman2024}%
  \BibitemOpen
  \bibfield  {author} {\bibinfo {author} {\bibfnamefont {J.}~\bibnamefont {Bierman}}, \bibinfo {author} {\bibfnamefont {Y.}~\bibnamefont {Li}},\ and\ \bibinfo {author} {\bibfnamefont {J.}~\bibnamefont {Lu}},\ }\bibfield  {title} {\bibinfo {title} {Qubit count reduction by orthogonally constrained orbital optimization for variational quantum excited-state solvers},\ }\href@noop {} {\bibfield  {journal} {\bibinfo  {journal} {Journal of Chemical Theory and Computation}\ }\textbf {\bibinfo {volume} {20}},\ \bibinfo {pages} {3131} (\bibinfo {year} {2024})}\BibitemShut {NoStop}%
\bibitem [{\citenamefont {Marie}\ and\ \citenamefont {Burton}(2023)}]{marie2023excited}%
  \BibitemOpen
  \bibfield  {author} {\bibinfo {author} {\bibfnamefont {A.}~\bibnamefont {Marie}}\ and\ \bibinfo {author} {\bibfnamefont {H.~G.}\ \bibnamefont {Burton}},\ }\bibfield  {title} {\bibinfo {title} {Excited states, symmetry breaking, and unphysical solutions in state-specific casscf theory},\ }\href@noop {} {\bibfield  {journal} {\bibinfo  {journal} {The Journal of Physical Chemistry A}\ }\textbf {\bibinfo {volume} {127}},\ \bibinfo {pages} {4538} (\bibinfo {year} {2023})}\BibitemShut {NoStop}%
\bibitem [{\citenamefont {Kossoski}\ and\ \citenamefont {Loos}(2023)}]{kossoski2023ssci}%
  \BibitemOpen
  \bibfield  {author} {\bibinfo {author} {\bibfnamefont {F.}~\bibnamefont {Kossoski}}\ and\ \bibinfo {author} {\bibfnamefont {P.-F.}\ \bibnamefont {Loos}},\ }\bibfield  {title} {\bibinfo {title} {State-specific configuration interaction for excited states},\ }\href {https://doi.org/10.1021/acs.jctc.3c00057} {\bibfield  {journal} {\bibinfo  {journal} {Journal of Chemical Theory and Computation}\ }\textbf {\bibinfo {volume} {19}},\ \bibinfo {pages} {2258} (\bibinfo {year} {2023})},\ \bibinfo {note} {pMID: 37024102},\ \Eprint {https://arxiv.org/abs/https://doi.org/10.1021/acs.jctc.3c00057} {https://doi.org/10.1021/acs.jctc.3c00057} \BibitemShut {NoStop}%
\bibitem [{\citenamefont {Yalouz}\ and\ \citenamefont {Robert}(2023)}]{yalouz2023ss}%
  \BibitemOpen
  \bibfield  {author} {\bibinfo {author} {\bibfnamefont {S.}~\bibnamefont {Yalouz}}\ and\ \bibinfo {author} {\bibfnamefont {V.}~\bibnamefont {Robert}},\ }\bibfield  {title} {\bibinfo {title} {Orthogonally constrained orbital optimization: Assessing changes of optimal orbitals for orthogonal multireference states},\ }\href {https://doi.org/10.1021/acs.jctc.2c01144} {\bibfield  {journal} {\bibinfo  {journal} {Journal of Chemical Theory and Computation}\ }\textbf {\bibinfo {volume} {19}},\ \bibinfo {pages} {1388} (\bibinfo {year} {2023})},\ \bibinfo {note} {pMID: 36790330},\ \Eprint {https://arxiv.org/abs/https://doi.org/10.1021/acs.jctc.2c01144} {https://doi.org/10.1021/acs.jctc.2c01144} \BibitemShut {NoStop}%
\bibitem [{\citenamefont {Saade}\ and\ \citenamefont {Burton}(2024)}]{saade2024excited}%
  \BibitemOpen
  \bibfield  {author} {\bibinfo {author} {\bibfnamefont {S.}~\bibnamefont {Saade}}\ and\ \bibinfo {author} {\bibfnamefont {H.~G.}\ \bibnamefont {Burton}},\ }\bibfield  {title} {\bibinfo {title} {Excited state-specific casscf theory for the torsion of ethylene},\ }\href@noop {} {\bibfield  {journal} {\bibinfo  {journal} {Journal of Chemical Theory and Computation}\ }\textbf {\bibinfo {volume} {20}},\ \bibinfo {pages} {5105} (\bibinfo {year} {2024})}\BibitemShut {NoStop}%
\bibitem [{\citenamefont {Burton}(2021)}]{burton2021generalized}%
  \BibitemOpen
  \bibfield  {author} {\bibinfo {author} {\bibfnamefont {H.~G.~A.}\ \bibnamefont {Burton}},\ }\bibfield  {title} {\bibinfo {title} {Generalized nonorthogonal matrix elements: Unifying wick’s theorem and the slater–condon rules},\ }\href {https://doi.org/10.1063/5.0045442} {\bibfield  {journal} {\bibinfo  {journal} {The Journal of Chemical Physics}\ }\textbf {\bibinfo {volume} {154}},\ \bibinfo {pages} {144109} (\bibinfo {year} {2021})},\ \Eprint {https://arxiv.org/abs/https://pubs.aip.org/aip/jcp/article-pdf/doi/10.1063/5.0045442/15587218/144109\_1\_online.pdf} {https://pubs.aip.org/aip/jcp/article-pdf/doi/10.1063/5.0045442/15587218/144109\_1\_online.pdf} \BibitemShut {NoStop}%
\bibitem [{\citenamefont {Burton}(2022)}]{burton2022generalized}%
  \BibitemOpen
  \bibfield  {author} {\bibinfo {author} {\bibfnamefont {H.~G.}\ \bibnamefont {Burton}},\ }\bibfield  {title} {\bibinfo {title} {Generalized nonorthogonal matrix elements. ii: Extension to arbitrary excitations},\ }\href@noop {} {\bibfield  {journal} {\bibinfo  {journal} {The Journal of Chemical Physics}\ }\textbf {\bibinfo {volume} {157}} (\bibinfo {year} {2022})}\BibitemShut {NoStop}%
\bibitem [{\citenamefont {Zhu}\ \emph {et~al.}(2025)\citenamefont {Zhu}, \citenamefont {Bierman}, \citenamefont {Lu},\ and\ \citenamefont {Li}}]{zhu2025}%
  \BibitemOpen
  \bibfield  {author} {\bibinfo {author} {\bibfnamefont {G.}~\bibnamefont {Zhu}}, \bibinfo {author} {\bibfnamefont {J.}~\bibnamefont {Bierman}}, \bibinfo {author} {\bibfnamefont {J.}~\bibnamefont {Lu}},\ and\ \bibinfo {author} {\bibfnamefont {Y.}~\bibnamefont {Li}},\ }\bibfield  {title} {\bibinfo {title} {Quantum circuit for non-unitary linear transformation of basis sets},\ }\href@noop {} {\bibfield  {journal} {\bibinfo  {journal} {arXiv preprint arXiv:2502.08962}\ } (\bibinfo {year} {2025})}\BibitemShut {NoStop}%
\bibitem [{\citenamefont {Parrish}\ \emph {et~al.}(2019)\citenamefont {Parrish}, \citenamefont {Hohenstein}, \citenamefont {McMahon},\ and\ \citenamefont {Mart\'{\i}nez}}]{parrish2019mcvqe}%
  \BibitemOpen
  \bibfield  {author} {\bibinfo {author} {\bibfnamefont {R.~M.}\ \bibnamefont {Parrish}}, \bibinfo {author} {\bibfnamefont {E.~G.}\ \bibnamefont {Hohenstein}}, \bibinfo {author} {\bibfnamefont {P.~L.}\ \bibnamefont {McMahon}},\ and\ \bibinfo {author} {\bibfnamefont {T.~J.}\ \bibnamefont {Mart\'{\i}nez}},\ }\bibfield  {title} {\bibinfo {title} {Quantum computation of electronic transitions using a variational quantum eigensolver},\ }\href {https://doi.org/10.1103/PhysRevLett.122.230401} {\bibfield  {journal} {\bibinfo  {journal} {Phys. Rev. Lett.}\ }\textbf {\bibinfo {volume} {122}},\ \bibinfo {pages} {230401} (\bibinfo {year} {2019})}\BibitemShut {NoStop}%
\bibitem [{\citenamefont {Nakanishi}\ \emph {et~al.}(2019)\citenamefont {Nakanishi}, \citenamefont {Mitarai},\ and\ \citenamefont {Fujii}}]{nakanishi2019ssvqe}%
  \BibitemOpen
  \bibfield  {author} {\bibinfo {author} {\bibfnamefont {K.~M.}\ \bibnamefont {Nakanishi}}, \bibinfo {author} {\bibfnamefont {K.}~\bibnamefont {Mitarai}},\ and\ \bibinfo {author} {\bibfnamefont {K.}~\bibnamefont {Fujii}},\ }\bibfield  {title} {\bibinfo {title} {Subspace-search variational quantum eigensolver for excited states},\ }\href {https://doi.org/10.1103/PhysRevResearch.1.033062} {\bibfield  {journal} {\bibinfo  {journal} {Phys. Rev. Res.}\ }\textbf {\bibinfo {volume} {1}},\ \bibinfo {pages} {033062} (\bibinfo {year} {2019})}\BibitemShut {NoStop}%
\bibitem [{\citenamefont {Higgott}\ \emph {et~al.}(2019)\citenamefont {Higgott}, \citenamefont {Wang},\ and\ \citenamefont {Brierley}}]{higgott2019variational}%
  \BibitemOpen
  \bibfield  {author} {\bibinfo {author} {\bibfnamefont {O.}~\bibnamefont {Higgott}}, \bibinfo {author} {\bibfnamefont {D.}~\bibnamefont {Wang}},\ and\ \bibinfo {author} {\bibfnamefont {S.}~\bibnamefont {Brierley}},\ }\bibfield  {title} {\bibinfo {title} {Variational quantum computation of excited states},\ }\href@noop {} {\bibfield  {journal} {\bibinfo  {journal} {Quantum}\ }\textbf {\bibinfo {volume} {3}},\ \bibinfo {pages} {156} (\bibinfo {year} {2019})}\BibitemShut {NoStop}%
\bibitem [{\citenamefont {Bierman}\ \emph {et~al.}(2022)\citenamefont {Bierman}, \citenamefont {Li},\ and\ \citenamefont {Lu}}]{bierman2022qomm}%
  \BibitemOpen
  \bibfield  {author} {\bibinfo {author} {\bibfnamefont {J.}~\bibnamefont {Bierman}}, \bibinfo {author} {\bibfnamefont {Y.}~\bibnamefont {Li}},\ and\ \bibinfo {author} {\bibfnamefont {J.}~\bibnamefont {Lu}},\ }\bibfield  {title} {\bibinfo {title} {Quantum {Orbital} {Minimization} {Method} for {Excited} {States} {Calculation} on a {Quantum} {Computer}},\ }\href {https://doi.org/10.1021/acs.jctc.2c00218} {\bibfield  {journal} {\bibinfo  {journal} {Journal of Chemical Theory and Computation}\ }\textbf {\bibinfo {volume} {18}},\ \bibinfo {pages} {4674} (\bibinfo {year} {2022})},\ \bibinfo {note} {\_eprint: https://doi.org/10.1021/acs.jctc.2c00218}\BibitemShut {NoStop}%
\bibitem [{\citenamefont {Powell}(2007)}]{powell2007view}%
  \BibitemOpen
  \bibfield  {author} {\bibinfo {author} {\bibfnamefont {M.~J.}\ \bibnamefont {Powell}},\ }\bibfield  {title} {\bibinfo {title} {A view of algorithms for optimization without derivatives},\ }\href@noop {} {\bibfield  {journal} {\bibinfo  {journal} {Mathematics Today-Bulletin of the Institute of Mathematics and its Applications}\ }\textbf {\bibinfo {volume} {43}},\ \bibinfo {pages} {170} (\bibinfo {year} {2007})}\BibitemShut {NoStop}%
\bibitem [{\citenamefont {Powell}(1994)}]{powell1994cobyla}%
  \BibitemOpen
  \bibfield  {author} {\bibinfo {author} {\bibfnamefont {M.~J.}\ \bibnamefont {Powell}},\ }\href@noop {} {\emph {\bibinfo {title} {A direct search optimization method that models the objective and constraint functions by linear interpolation}}}\ (\bibinfo  {publisher} {Springer},\ \bibinfo {year} {1994})\BibitemShut {NoStop}%
\bibitem [{\citenamefont {Kraft}(1988)}]{kraft1988slsqp}%
  \BibitemOpen
  \bibfield  {author} {\bibinfo {author} {\bibfnamefont {D.}~\bibnamefont {Kraft}},\ }\href@noop {} {\emph {\bibinfo {title} {A software package for sequential quadratic programming}}},\ \bibinfo {type} {Tech. Rep.}\ \bibinfo {number} {DFVLR-FB 88-28}\ (\bibinfo  {institution} {Deutsche Forschungs- und Versuchsanstalt f{\"u}r Luft- und Raumfahrt},\ \bibinfo {year} {1988})\BibitemShut {NoStop}%
\bibitem [{\citenamefont {Liu}\ and\ \citenamefont {Li}(2025)}]{liu2025upoqa}%
  \BibitemOpen
  \bibfield  {author} {\bibinfo {author} {\bibfnamefont {Y.}~\bibnamefont {Liu}}\ and\ \bibinfo {author} {\bibfnamefont {Y.}~\bibnamefont {Li}},\ }\bibfield  {title} {\bibinfo {title} {A model-based derivative-free optimization algorithm for partially separable problems},\ }\href@noop {} {\bibfield  {journal} {\bibinfo  {journal} {arXiv preprint arXiv:2506.21948}\ } (\bibinfo {year} {2025})}\BibitemShut {NoStop}%
\end{thebibliography}%

\end{document}